\documentclass[11pt,a4paper]{article}
\usepackage[most]{tcolorbox}
\usepackage{jcappub}
\usepackage{graphicx}
\usepackage{amsmath}
\usepackage{epstopdf}
\usepackage{latexsym}
\usepackage{dcolumn}
\usepackage{bm}
\usepackage{url}
\usepackage{hyperref}

\input{epsf}
\newcommand{\be}{\begin{equation}}
\newcommand{\ee}{\end{equation}}
\newcommand{\ba}{\begin{eqnarray}}
\newcommand{\ea}{\end{eqnarray}}
\newcommand{\bi}{\begin{itemize}}
\newcommand{\ei}{\end{itemize}}
\newcommand{\ga}{\gtrsim}
\newcommand{\bfi}{\begin{figure}
\epsfxsize=9cm
\epsffile}
\newcommand{\bfinew}{\begin{figure}
\begin{center}
\includegraphics}
\newcommand{\efi}{\end{figure}}
\newcommand{\efinew}{
\end{center}
\end{figure}}

\newcommand{\la}{\lesssim}

\title{Baryonic feedback suppression of the matter power spectrum: a three-parameter fitting formula and its single-parameter reduction} 
\author[a,b,c,d]{Pengjie Zhang}
\affiliation[a]{School of Physics and Astronomy, Shanghai Jiao Tong University, Shanghai 200240, China}
\affiliation[b]{Tsung-Dao Lee Institute, Shanghai Jiao Tong University, Shanghai 201210, China}
\affiliation[c]{State Key Laboratory of Dark Matter Physics, Shanghai 200240, China}
\affiliation[d]{Key Laboratory for Particle Astrophysics and Cosmology (MOE)/Shanghai Key Laboratory for Particle Physics and Cosmology, Shanghai 200240, China}

\emailAdd{zhangpj@sjtu.edu.cn}

\abstract{Baryonic feedback can suppress matter clustering by $\sim 10\%$ at $k\sim 1h$/Mpc and $z\sim 1$, and is therefore a major source of systematic errors in weak lensing cosmology. We investigate this effect through the ratio $S(k,z)\equiv P_{\rm hydro}(k,z)/P_{\rm DMO}(k,z)$ measured in the
FLAMINGO, IllustrisTNG, and Illustris hydrodynamical simulation suites, together with the one-parameter-at-a-time (1P) variation runs of CAMELS-TNG and CAMELS-SIMBA.  We present a three-parameter fitting formula that reproduces all but one of these runs with maximum error less than $0.03$ (and typical error less than $0.01$) at $k\leq 3h/$Mpc and $z\leq 2$; the only exception is a run with unrealistically low $\Omega_m=0.1$. Each of the three parameters ($B_0$, $n$, and $a$) has a clear physical interpretation: $B_0$ sets the characteristic scale of the suppression at $z=0$, corresponding to a gas particle displacement scale of $B_0^{1/2}$; $n$ sets the suppression floor $(1-\Omega_b/\Omega_m)^n$; and $a$ sets the redshift $z_B$ at which the suppression peaks. Furthermore, we find that $B_0$ and $z_B$ are significantly correlated for most runs. Also, $n\simeq 1$ for FLAMINGO, TNG and CAMELS-TNG. For these simulations, the description reduces to a single parameter with maximum error less than $0.03$.  The formulas remain accurate at max$(|\Delta S|)<0.03$ (and typical error less than $0.01$) for cosmologies well away from the calibration cosmology, spanning the range $\Omega_m\in [0.2,0.5]$, $\Omega_b\in [0.029,0.069]$ and $\sigma_8\in [0.6,1.0]$. These results imply that, despite the diversity of baryonic physics implementations, an accurate description of $S(k,z)$ requires at most three effective degrees of freedom. Therefore the impact of feedback on weak lensing cosmology can in principle be  mitigated  internally without significant loss of  cosmological constraining power. }

\begin{document}
\maketitle
\section{Introduction}
The matter power spectrum  $P_m(k)$ plays a central role in describing 
the large scale structure of the universe, and therefore in probing cosmological physics such as the nature of dark energy \cite{2013PhR...530...87W,2019ARA&A..57..335F,2019LRR....22....1I,2026RAA....26h4006Z,2026RAA....26h4001R}. At leading order, the evolution of both dark matter and baryons is determined by the same gravitational physics. So the corresponding power spectrum can be computed robustly through  dark matter only (DMO)  simulations and emulators based on these simulations (e.g., the Coyote project \cite{2009ApJ...705..156H}; the Aemulus emulator \cite{2019ApJ...875...69D}; the BACCO emulator \cite{2021MNRAS.507.5869A}; the EuclidEmulator and EuclidEmulator2 \cite{2021MNRAS.505.2840E}; the Mira-Titan emulator \cite{2023MNRAS.520.3443M}; the CSST emulator \cite{2025SCPMA..6889512C}; Dark Quest and Dark Quest2 emulators \cite{2026arXiv260528596T}). For example, the recently constructed CSST emulator \cite{2025SCPMA..6889512C,2025SCPMA..6809513C,2025SCPMA..6829512Z} is able to predict the DMO power spectrum $P^{\rm DMO}_m(k)$ with better than $1\%$ accuracy to $k=10h/$Mpc. This emulator is based upon 129  simulations, each with $3072^3$  particles,  sampling an 8-dimensional  cosmological parameter hyperspace. Furthermore, its extended form  \cite{2026JCAP...06..086C} covers the dark energy parameter space of DESI \cite{2025PhRvD.112h3515A}.  

However, the baryon distribution and evolution are inevitably affected by non-gravitational mechanisms such as gas cooling, star formation, supernova (SN) and AGN feedback (e.g., \cite{2024MNRAS.528.3797W,2025A&A...697A..63O}). This non-gravitational/baryonic effect has been recognized as   a major systematic error of weak lensing cosmology (e.g., \cite{2020JCAP...04..019S,2021A&A...649A.100M,2024MNRAS.527.5206Y}). The baryonic effect on the matter power spectrum and weak lensing cosmology  was first pointed out in theory by \cite{2004APh....22..211W,2004ApJ...616L..75Z}, then heavily investigated by hydrodynamical simulations \cite{2006ApJ...640L.119J,2008ApJ...672...19R,2011MNRAS.415.3649V,2014MNRAS.444.1518V,2018MNRAS.475..676S,2019MNRAS.486.2827D,2020MNRAS.491.2424V,2023MNRAS.526.4978S,2023MNRAS.526.6103K,2026A&C....5701159H,CAMELS_presentation,CAMELS_DR1,CAMELS_DR2,CAMELS_DR3}. At scale $k\sim 1h/$Mpc most relevant for weak lensing cosmology, feedback by AGN and SN dominates over gas cooling and results in a $\sim 1\%$ to $\sim 10\%$ suppression of the matter power spectrum at $z=0$, with respect to the DMO case. Despite the inconclusiveness of hydro simulations, many recent observational analyses support significant suppression  of $\sim 10\%$. There exist multiple lines of independent evidence from the thermal Sunyaev-Zel'dovich (tSZ) effect \cite{2023ApJ...953..188C,2026JCAP...03..036D}, the kinematic SZ (kSZ) effect \cite{2025PhRvD.112h3509H,2026MNRAS.550g1314B},  X-ray \cite{2024MNRAS.528.4379G},  weak lensing \cite{2023MNRAS.518.5340C,2024PhRvD.110f3532X} and joint analysis of various data combinations \cite{2022MNRAS.514.3802S, 2024MNRAS.534..655B,2026MNRAS.549ag993S}.  

Therefore a key task in weak lensing cosmology is to model the baryonic feedback effect accurately. 
For the weak lensing angular power spectrum, a typical angular scale $\ell \sim 10^3$ corresponds to $k\sim 1 h/$Mpc at the lens redshift $z_{\rm L}\sim 0.5$ most relevant for  typical source redshift $z_{\rm s}\sim 1$. Therefore, this sets the requirement of modeling the matter clustering  accurately at $k\la 3h/$Mpc and $z\la 2$. Besides direct efforts of hydro simulations and simulation based emulators (e.g., the BACCO emulator \cite{2021MNRAS.506.4070A}; the CAMELS emulator \cite{2025MNRAS.538.1415S} and the FLAMINGO emulator \cite{2025MNRAS.539.1337S}), there exist analytical and semi-analytical models, such as the halo model based HMCODE \cite{2015MNRAS.454.1958M,2021MNRAS.502.1401M} and analytical  fitting formulas \cite{2020MNRAS.491.2424V,2023MNRAS.523.2247S,2025MNRAS.540.2322S}. On the other hand, the baryonification method displaces particles in N-body simulations with analytical recipes to model various baryonic effects \cite{2015JCAP...12..049S,2021MNRAS.503.3596A,2025JCAP...12..043S,2025JCAP...11..046K}. 

\bfinew[width=1.0\textwidth]{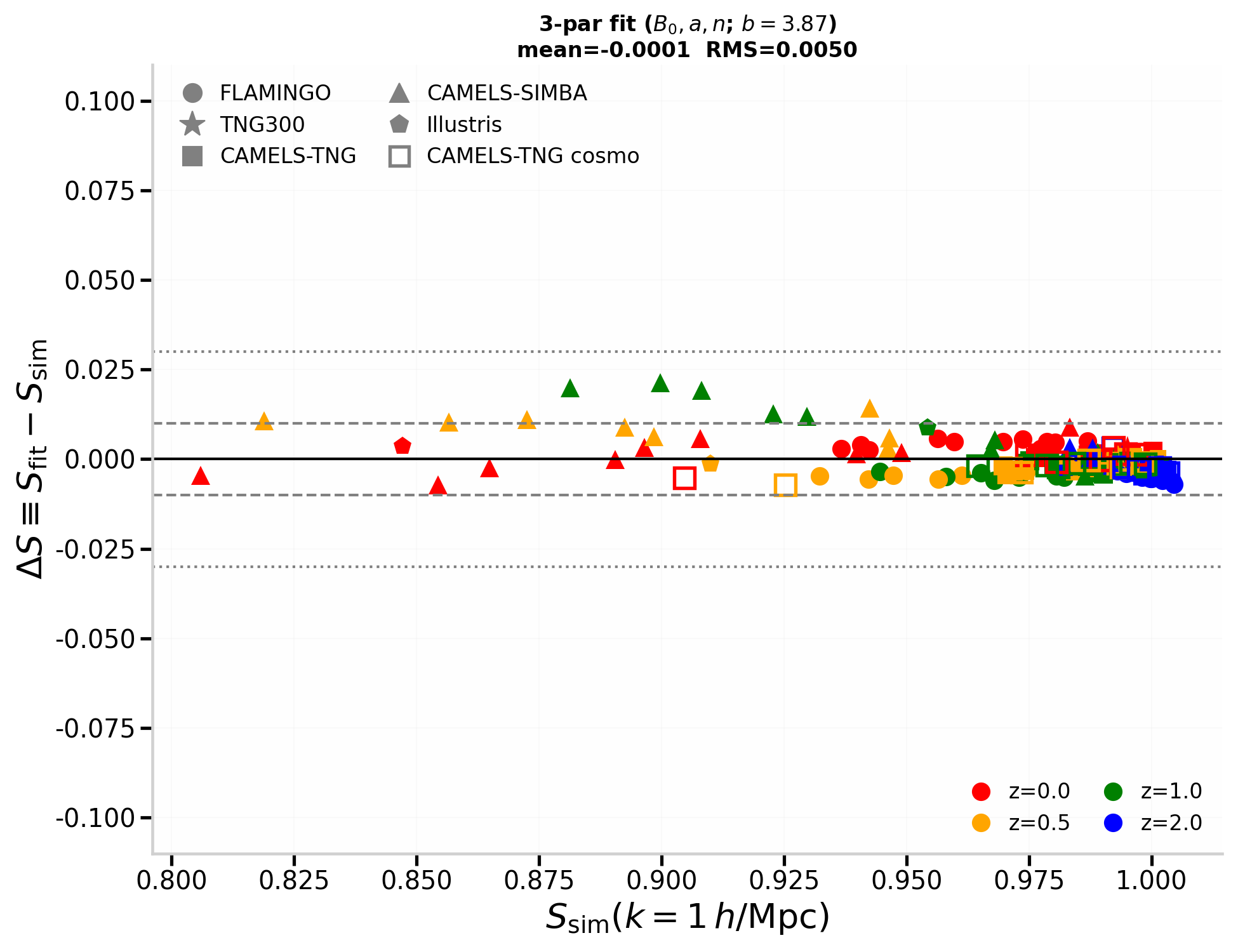
}
\caption{Performance of the three-parameter fitting formula at $k=1h/$Mpc. It is accurate to max$(|\Delta S|)<0.02$ for a variety of baryonic physics and cosmology. 
\label{fig:3Perrork1}}
\efinew

Taking advantage of accurate first-principle prediction of the DMO power spectrum, an efficient strategy is to model the  suppression function $S(k,z)$, 
\begin{eqnarray}
S(k,z)\equiv \frac{P_{m}(k,z)}{P_m^{\rm DMO}(k,z)}\ . 
\end{eqnarray}
The modeling of $S(k,z)$ must on one hand be sufficiently versatile in amplitude, shape and redshift evolution to incorporate a variety of baryonic physics, and on the other hand must contain as few free parameters as possible. From the viewpoint of hydro simulations, there are many subgrid physics parameters (e.g., 30 considered in CAMELS-TNG runs). At least  four standard parameters describing AGN and SN feedback are directly relevant (e.g., in IllustrisTNG \cite{2018MNRAS.475..676S} and FLAMINGO  \cite{2023MNRAS.526.4978S}).  From the viewpoint of analytical and semi-analytical modeling, there are also many parameters to be considered. For example, for the halo model approach,  in general 4 or more parameters are needed to specify the gas fraction and distribution as a function of halo mass and redshift.  However, in terms of $S(k,z)$, consequences of these parameters may be degenerate.  

\bfinew[width=1.0\textwidth]{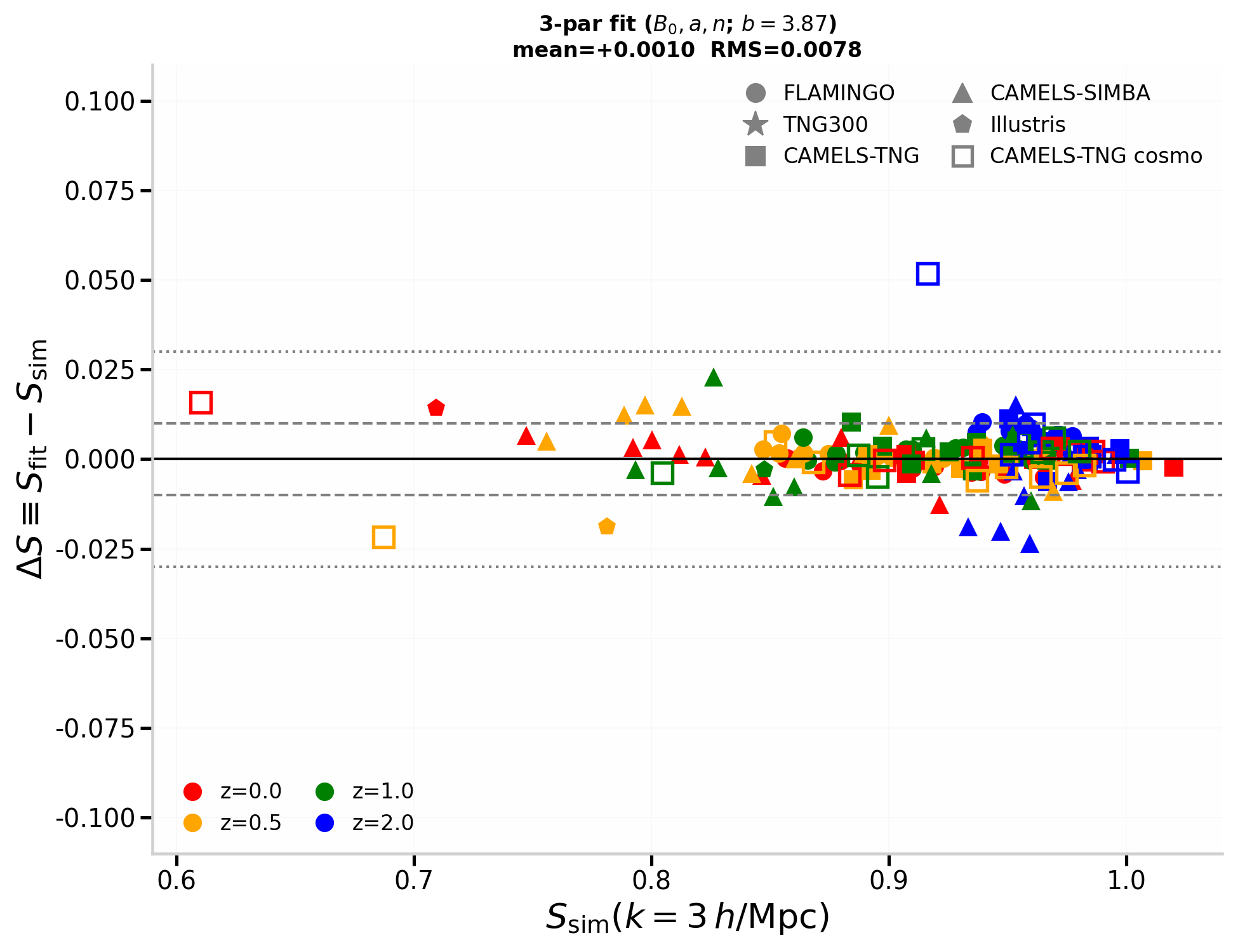
}
\caption{The three-parameter fitting formula remains accurate at $k=3h/$Mpc for a variety of baryonic physics and cosmology. The largest error occurs for the CAMELS-TNG cosmology variation run 1P\_p1\_n2. It has  $\Omega_m=0.1$ and therefore unrealistically high $49\%$ baryon fraction, included only for the purpose of testing the limit of our fitting formula. \label{fig:3Perrork3}}
\efinew

Then how many effective degrees of freedom (DoFs) are required to model $S(k,z)$ accurately for weak lensing cosmology? We argue that there are at most three effective DoFs, as we find a three-parameter fitting formula with max$(|\Delta S|)\la 0.03$ for FLAMINGO, TNG300, CAMELS-TNG variations, Illustris and CAMELS-SIMBA variations of vastly different feedback implementations.  Surprisingly,  the number of effective DoFs can be further reduced. For FLAMINGO and TNG simulations, it  reduces to $1$, demonstrated by the single-parameter fitting formula with max$(|\Delta S|)< 0.03$.  Even better, the three-parameter and single-parameter fitting formulas are robust to cosmology variations. They are trained under the cosmology of $\Omega_m\sim 0.3$, $\Omega_b\sim 0.05$ and $\sigma_8\sim 0.8$. But they remain accurate for significantly different cosmology such as $\Omega_m=0.2$, or $\Omega_b=0.069$, or $\sigma_8=1.0$.  Given the limited DoFs in $S(k,z)$, we expect that a weak lensing alone tomographic analysis is in principle able to disentangle the baryonic feedback from cosmology, without significant loss of cosmological constraining power. This is an issue to be further quantified in a companion paper. 

This paper is organized as follows. We first summarize the main results in \S \ref{sec:mainresults}. We present the basic fitting scheme and simulations for training and testing in \S \ref{sec:scheme}. Then we elaborate on technical details and describe the single-parameter fitting formula in \S \ref{sec:1P}. To incorporate the newly found effective DoFs in \S \ref{sec:1P} to improve the fitting, we extend to the three-parameter formula in \S \ref{sec:3P}. We present further discussion in \S \ref{sec:discussion}. 

\section{Summary of the main results}
\label{sec:mainresults}

We first present the three-parameter and single-parameter fitting formulas here. 
\begin{tcolorbox}[
    enhanced,
    colback=blue!5!white,        
    colframe=blue!60!black,     
    coltitle=white,
    colbacktitle=blue!60!black,  
    title={\textbf{The three-parameter and single-parameter formulas for the feedback suppression}},
    fonttitle=\small,
    boxrule=0.8pt,
    arc=2mm,
    left=2mm, right=2mm, top=1.5mm, bottom=1.5mm
]

\begin{equation}
\label{eqn:Sfit}
    S(k,z)=\frac{1+f_{\rm dm}^n B(z)k^2}{1+B(z)k^2}\ ,\  {\rm with}\ f_{\rm dm}=1-f_b\ ,\ f_b\equiv \frac{\Omega_b}{\Omega_m}\ .
\end{equation}
\begin{equation}
\label{eqn:B}
    B(z)=B_0(1+z)^{a}\exp(-bz)\ .
\end{equation}
\begin{itemize}
\item  The \textbf{three-parameter fitting formula}.  
\begin{eqnarray}
    {\rm varying}\ (B_0,a,n)\ .\  b=3.87\ .
    \end{eqnarray}
\item The \textbf{single-parameter fitting formula}. 
\begin{equation}
\label{eqn:1P}
 {\rm varying}\ B_0\ .\ n=1, b=3.87,c_0=1.03\ ,\ c_1=-0.20\ .
\end{equation}
\begin{equation}
z_B\equiv \frac{a}{b}-1=c_0+c_1\ln \left(1+B_0/10^{-2}\right)\ .
\end{equation}
\end{itemize}
Throughout this paper, $k$ is in unit of $h/$Mpc, and $B_0$ (and $B$) is in unit of $({\rm Mpc}/h)^2$. 
\end{tcolorbox}
$B_0$, $n$ and $z_B$ (or equivalently $a$) are the three effective degrees of freedom (DoFs) that we identify to accurately describe the feedback suppression. $l_0\equiv B^{1/2}_0$ corresponds to the feedback induced typical displacement of gas particles at $z=0$.  $n$  sets the maximum suppression depth as $1-f_{\rm dm}^n\simeq n\Omega_b/\Omega_m$.  $z_B$, the redshift of the peak suppression, quantifies the evolution of the suppression. 

Fig. \ref{fig:3Perrork1}, \ref{fig:3Perrork3}, \ref{fig:DS1} \& \ref{fig:DS2}
show the performance of the proposed formulas at two scales ($k=1h/$Mpc and $k=3h/$Mpc) and four redshifts ($z=0,0.5,1,2$) of 40 simulation runs. Table \ref{tab:errors} summarizes the performance in terms of the maximum absolute error (max$(|\Delta S|)$). In general, the single-parameter formula is accurate to max$(|\Delta S|)<0.03$ for FLAMINGO and TNG, but becomes inaccurate at $k>1h/$Mpc for strong feedback runs of Illustris and CAMELS-SIMBA. The three-parameter formula significantly enlarges the range of applicability. It is accurate to max$(|\Delta S|)<0.01$ ($0.03$) for moderate (strong) feedback runs. The above performance essentially persists for other cosmologies, except the $\Omega_m=0.1$ case that is already excluded by observations. Its failure arises from the unrealistically large baryon fraction $\Omega_b/\Omega_m=0.49$. 

\begin{table}[t]
\centering
\small
\setlength{\tabcolsep}{4pt}
\begin{tabular}{l|cc|cc|cc|cc}
\hline
\hline
Maximum error in $S(k,z)$ & \multicolumn{4}{c|}{Feedback variations} & \multicolumn{4}{c}{Cosmology variations} \\
\cline{2-9}
 & \multicolumn{2}{c|}{Moderate} & \multicolumn{2}{c|}{Strong} & \multicolumn{2}{c|}{$\Omega_m \geq 0.2$} & \multicolumn{2}{c}{\color{red}{$\Omega_m =0.1$}} \\
\cline{2-9}
 & $k = 1$ & $k = 3$ & $k = 1$ & $k = 3$ & $k = 1$ & $k = 3$ & $k = 1$ & $k = 3$ \\
\hline
Single-parameter   & $< 0.01$ & $<0.03$ & $< 0.04$ & \color{red}{$<0.13$}      & $<0.01$ & $<0.03$       & \color{red}{$<0.04$} & \color{red}{$<0.09$} \\
Three-parameter & $< 0.01$ & $< 0.01$ & $< 0.02$  & $< 0.03$      & $< 0.01$ & $<0.02$  & $< 0.01$ & \color{red}{$< 0.05$ }\\
\hline
\hline
\end{tabular}
\caption{Performance of the single-parameter
and three-parameter fitting formulas at $k = 1$ and $k = 3$ ($k$ in unit of
$h/$Mpc), for $z \leq 2$. For feedback variations, moderate refers to the
FLAMINGO, TNG300 and CAMELS-TNG 1P runs, and strong refers to the Illustris
and CAMELS-SIMBA 1P runs. For cosmology variations (the CAMELS-TNG 1P
cosmology runs with $\Omega_m \in \{0.1, 0.2, 0.4, 0.5\}$, $\sigma_8 \in
\{0.6, 1.0\}$, $\Omega_b \in \{0.029, 0.069\}$), the $\Omega_m = 0.1$ ($\Omega_b/\Omega_m=0.49$) case
serves as a stress test well beyond current observational constraints.  }
\label{tab:errors}
\end{table}

\bfinew[width=1.0\textwidth]{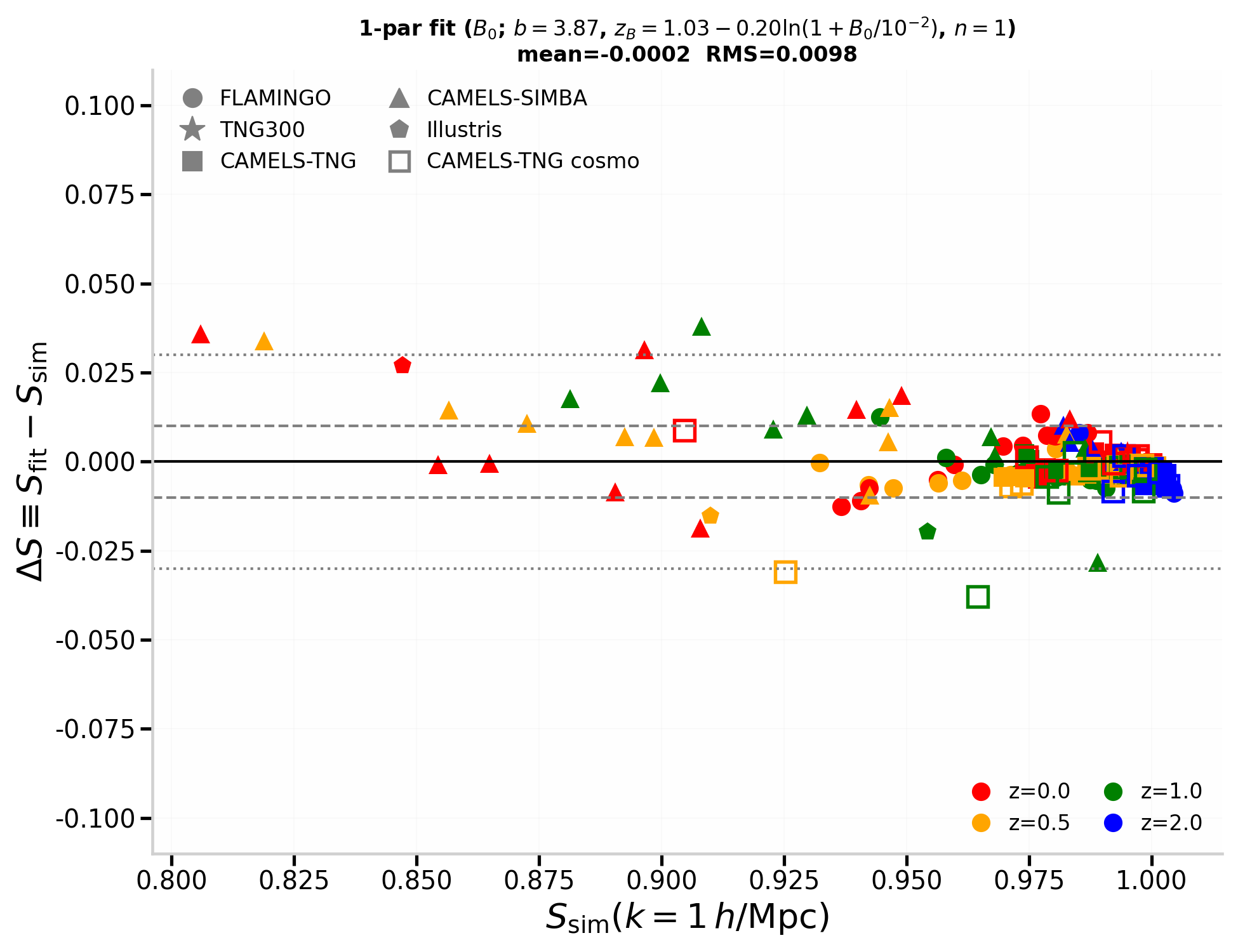}
\caption{Performance of the single-parameter fitting formula at $k=1h/$Mpc. The fitting accuracy up to this scale is slightly worse than, but comparable to the three-parameter fitting. This implies that there is approximately one effective degree of freedom to describe $S(k,z)$ at $k\leq 1h/$Mpc.  \label{fig:DS1}}
\efinew

\section{The parameterization scheme and the simulation test sets}
\label{sec:scheme}
\subsection{The parameterization of $S(k)$}
$S(k)$ should satisfy the following three conditions.
\begin{itemize}
    \item $S\rightarrow 1$ when $k\rightarrow 0$, since the weak equivalence principle guarantees that both baryons and dark matter follow the geodesic at sufficiently large scales where non-gravitational forces are negligible. 
    \item $\partial S/\partial k \rightarrow 0$ when $k\rightarrow 0$. This arises from the symmetry that $S(-{\bf k})=S({\bf k})$ where we restore the full dependence of $S$ on the wavevector ${\bf k}$. 
    \item $S\rightarrow f_{\rm dm}^n=(1-\Omega_b/\Omega_m)^n$ when $k\rightarrow \infty$. $n=0$ corresponds to the limit of no feedback ($S=1$). $n=2$ corresponds to the extreme case that feedback essentially eliminates baryon clustering below a certain scale. If we further consider the induced suppression in the later structure  growth, $n$ can be greater than $2$.  
\end{itemize}
\bfinew[width=1.0\textwidth]
{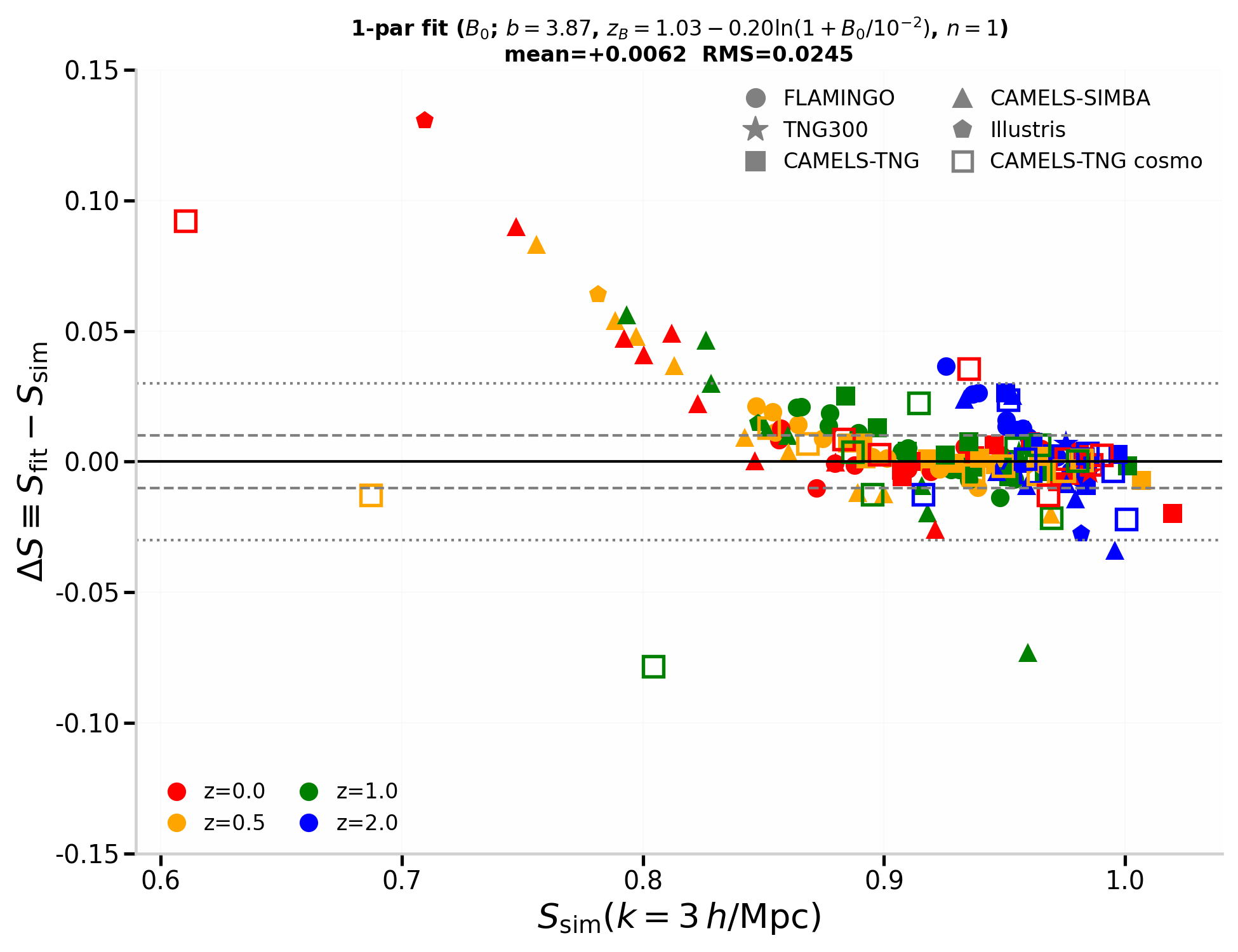}
\caption{The performance of the single-parameter ($B_0$) fitting formula at $k=3h/$Mpc. Typical  fitting accuracy for FLAMINGO and TNG series remains at $|\Delta S|\sim 0.01$. However, errors in Illustris and some of CAMELS-SIMBA degrade dramatically. This implies the existence of DoFs beyond $B_0$ at $k>1h/$Mpc. Three-parameter fitting formula captures two extra DoFs and pushes the applicability of the $k$ range to at least $3h/$Mpc. The single-parameter fitting formula also shows generality  over cosmological variations, except for the CAMELS-TNG cosmology variation run 1P\_p1\_n2 with unrealistically low $\Omega_m=0.1$.    \label{fig:DS2}}
\efinew

A simple form of $S(k)$ satisfying the above three requirements is Eq. \ref{eqn:Sfit}, in which the impact of feedback is compressed into the function $B(z)$ and $n$. To further understand their physical meaning, we consider a more physically motivated fitting scheme and then compare the two. Feedback displaces gas particles by ${\bf d}$ and therefore smooths the gas distribution by a window function $W(k)=\langle \exp(i{\bf k}\cdot{\bf d})\rangle=1-k^2\langle d^2\rangle/6+\cdots$.  Given that the baryon density is much smaller than the dark matter density, we may approximate the dark matter distribution as unperturbed. Then we have $\delta_m=(1-f_b)\delta_{\rm DM}+f_b\delta_{b}\simeq [(1-f_b)+f_bW]\delta_{\rm DMO}$. $S\simeq [(1-f_b)+f_bW]^2\rightarrow 1-f_bk^2\langle d^2\rangle/3$ when $k\langle d^2\rangle^{1/2}\ll 1$. Comparing this $S(k)$ to $S(k)$ of Eq. \ref{eqn:Sfit} when $kB^{1/2}\ll 1$, we find
\begin{equation}
    nB\simeq \langle d^2\rangle/3\ .
\end{equation}
Therefore $l_B\equiv B^{1/2}\sim \langle d^2\rangle^{1/2}$, up to a factor ($(3n)^{-1/2}$) of order unity. So $l_B$ denotes the typical displacement of gas particles by feedback. $B$ (or equivalently $l_B$) is a measure of  the feedback strength. 

On the other hand,  $n$ sets the minimum of $S(k)$, namely the suppression floor, as 
\begin{equation}
    S_{\rm floor}=f_{\rm dm}^n\simeq 1-n(\Omega_b/\Omega_m)\ .
\end{equation}
 It turns out later that this $f_{\rm dm}^n$ term captures most cosmological dependence and makes the fitting scheme applicable to cosmologies beyond the calibration cosmology. At $kl_B\ll 1$, $S\simeq 1-(1-f_{\rm dm}^n)Bk^2$. So $(1-f_{\rm dm}^n)B\simeq nB(\Omega_b/\Omega_m)$ controls the shape of the feedback suppression at large scale. This functional form of Eq. \ref{eqn:Sfit} is versatile. It is even able to describe the enhancement $S>1$ in the case of the  CAMELS-TNG 1P\_p3\_2 run, but with $n<0$. 

\subsection{Simulations to  test the fitting formulas}
Given large uncertainties in the feedback physics and numerical implementation, a useful fitting formula must be sufficiently robust to various feedback scenarios. We collect 32 simulation runs covering different  feedback strength, different AGN feedback mode, and different numerical implementations.

\bfinew[width=1.0\textwidth]{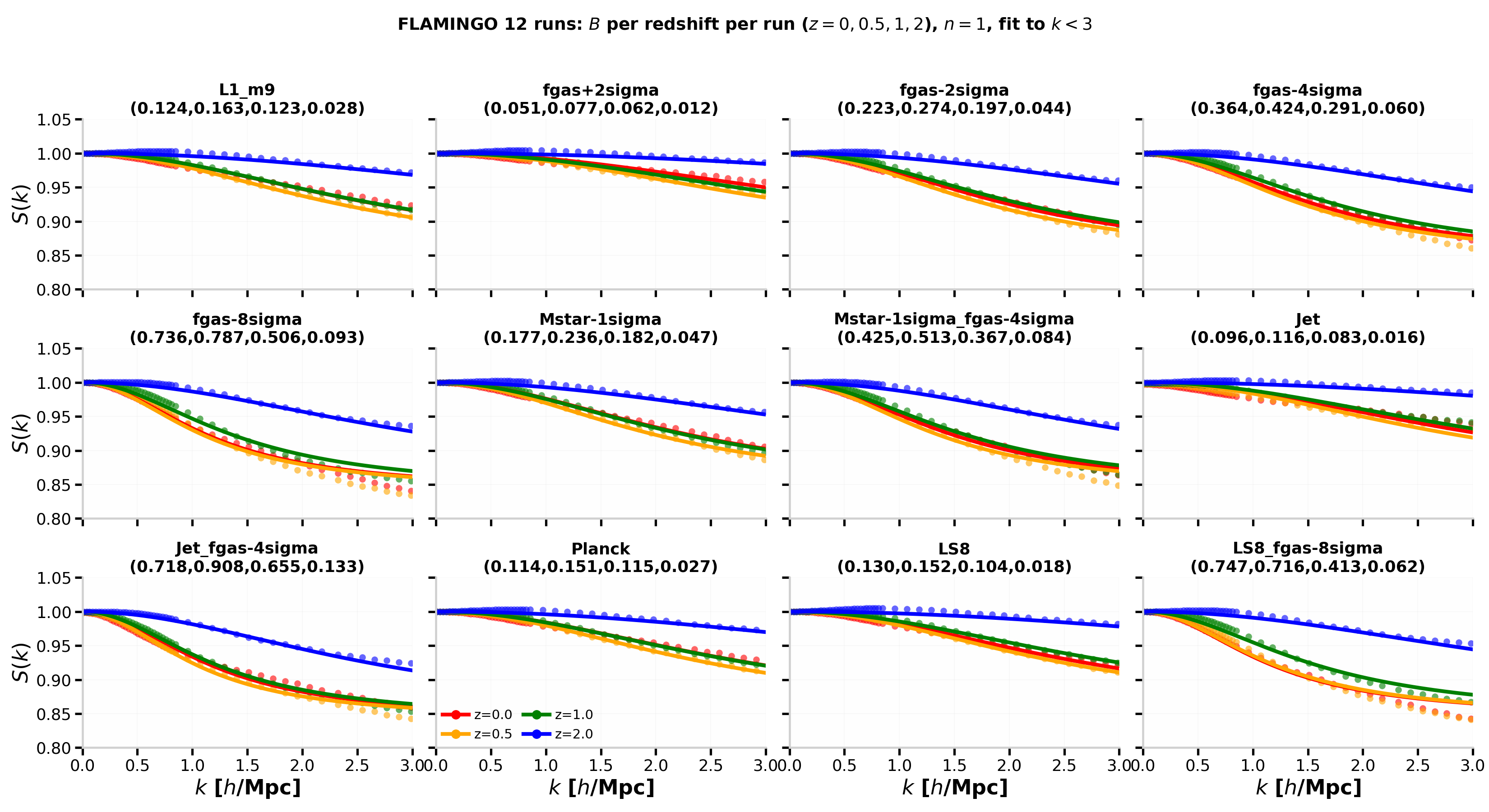}
\caption{Fitting the FLAMINGO series using Eq. \ref{eqn:Sfit} with $n=1$, but $B$ per redshift per run.   For clarity, we only show one third of  the data points. Each panel lists the value of $B$ at redshifts $z=0.0,0.5,1.0,2.0$. Eq. \ref{eqn:Sfit} works excellently.\label{fig:FLAMINGO}}
\efinew

\begin{itemize}
\item The moderate feedback series (in total 22 runs). Note that there is an exception, the CAMELS-TNG 1P\_p3\_2 run, whose feedback is too weak to suppress the power spectrum ($S(k)>1$).  Nonetheless, we include it in this series. 
\begin{itemize}
    \item The FLAMINGO series \cite{2023MNRAS.526.4978S,2023MNRAS.526.6103K,2026A&C....5701159H}. We choose 12 runs, including L1\_m9, fgas$\pm 2\sigma$, fgas-4$\sigma$, fgas-8$\sigma$, M$^*$-$\sigma$, M$^*$-$\sigma$\_fgas-4$\sigma$, Jet, Jet\_fgas-4$\sigma$, Planck, LS8 and LS8\_fgas-8$\sigma$. We directly use the power spectra provided through the website\footnote{\url{https://dataweb.cosma.dur.ac.uk:8443/flamingo/index.html}} and convert the unit of $k$ from $1/$Mpc to $h/$Mpc. 
     \item The IllustrisTNG \cite{2018MNRAS.475..676S} and CAMELS-TNG series \cite{CAMELS_presentation,CAMELS_DR1,CAMELS_DR2,CAMELS_DR3}. Within the TNG series, we choose TNG300.\footnote{The author thanks Shuren Zhou  for providing the TNG300 and Illustris-1 power spectra that he calculated using  the simulation snapshots.} Within the CAMELS-TNG runs, we choose the 1P set with $50\ {\rm Mpc}/h$ boxsize.  Within the 1P set, we only consider the fiducial run 1P\_p1\_0 and  8 runs with identifiers 1P\_pX\_Y (X=$3,4,5,6$. Y$=2$, n2). These 8 runs vary the 4 standard IllustrisTNG baryonic feedback parameters ($X=3,4,5,6=A_{\rm SN1}, A_{\rm AGN1}$, $A_{\rm SN2}$, $A_{\rm AGN2}$). $A_{\rm SN1}$ is the energy per unit star formation rate of the galactic winds. $A_{\rm AGN1}$ is the energy per unit black hole accretion rate. $A_{\rm SN2}$ is the wind speed of the galactic winds. $A_{\rm AGN2}$ is the ejection speed/burstiness of the kinetic mode of the black hole feedback. We directly use the power spectra provided in the CAMELS website.\footnote{\url{https://users.flatironinstitute.org/~camels/Pk/IllustrisTNG/L50n512/1P/} and the DMO counterpart.}  Note that within the 1P set,  each run only varies a single parameter denoted by X. In total we choose 10 runs. 
\end{itemize}
    \item The strong feedback series (in total 10 runs). Besides the Illustris simulation \cite{2014MNRAS.444.1518V}, we also include 9 of the  1P set of the CAMELS-SIMBA runs \cite{CAMELS_presentation,CAMELS_DR1,CAMELS_DR2,CAMELS_DR3}, which  have relatively small boxsize $25\ {\rm Mpc}/h$. We only consider the runs with the fiducial cosmology ($\Omega_m=0.3,\sigma_8=0.8, \Omega_b=0.049$). Besides the fiducial run, we consider 8 runs by varying the 4 standard SIMBA baryonic feedback parameters. They are $A_{\rm SN1}$ for the mass loading of the galactic winds, $A_{\rm AGN1}$ for the momentum flux of the QSO \& jet-mode black-hole feedback, $A_{\rm SN2}$ for the wind speed of the galactic winds, and $A_{\rm AGN2}$ for the jet speed of the jet-mode black-hole feedback. Again, we directly use the power spectra provided by the CAMELS website.\footnote{\url{https://users.flatironinstitute.org/~camels/Pk/SIMBA/L25n256/1P/} and the DMO counterpart.} Note that some but not all runs in this series have strong feedback. For example, the feedback strength of the CAMELS-SIMBA 1P\_p5\_2 run is stronger than all TNG runs, but weaker than the FLAMINGO Jet\_fgas-4sigma run. Nevertheless, the feedback strength of  Illustris and CAMELS-SIMBA 1P\_p6\_2 is the strongest among all feedback variation runs. 
\end{itemize}
For brevity we investigate only four redshifts ($z=0,0.5,1.0,2$).
These simulation runs have vastly different hydro solvers, baryonic physics and algorithms of subgrid physics. So they are suitable for testing the generality of the proposed fitting formula against these complexities.

\bfinew[width=1.0\textwidth]{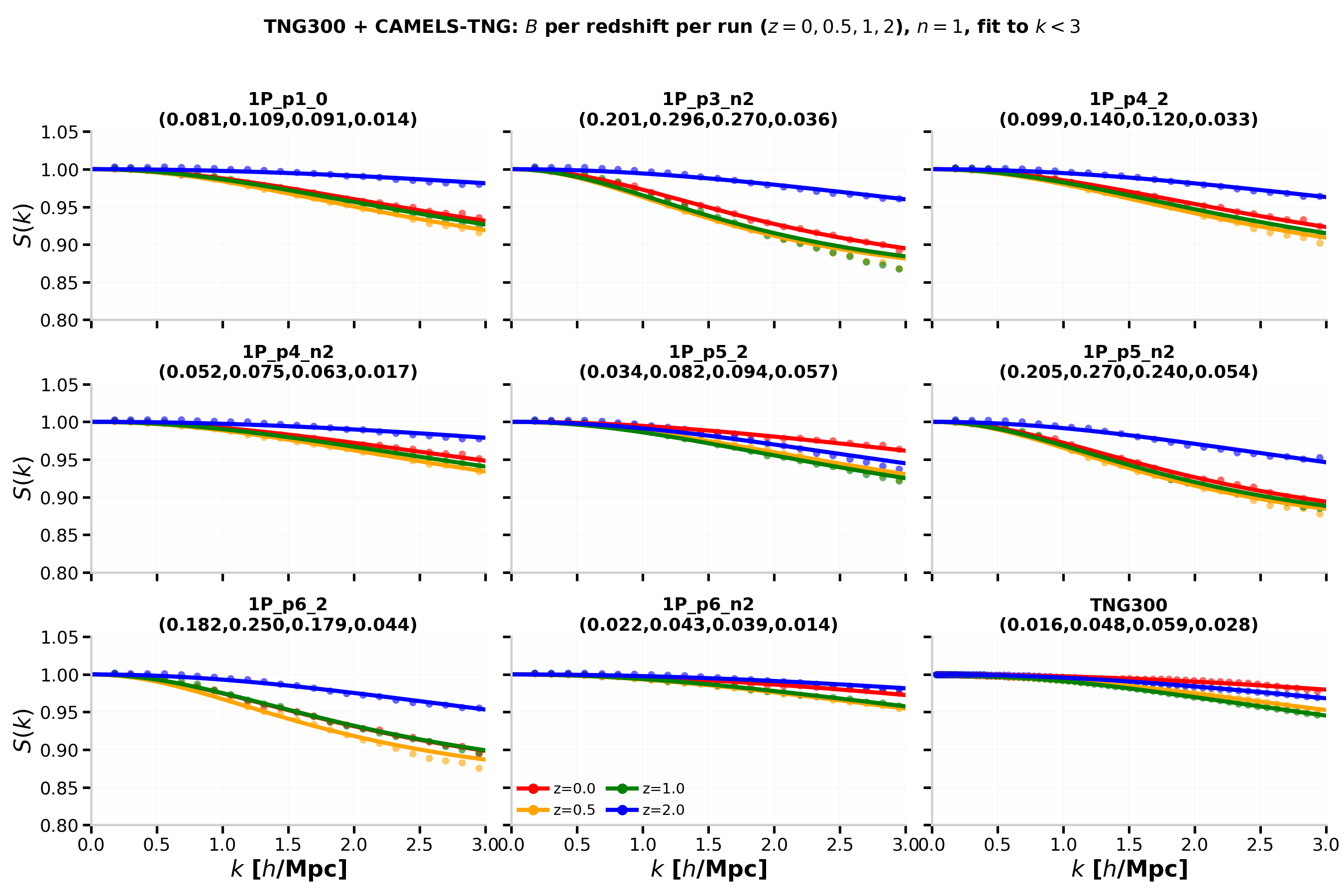}
\caption{The same as Fig. \ref{fig:FLAMINGO}, but for the TNG series.  The fitting formula (Eq. \ref{eqn:Sfit}) works excellently.  \label{fig:TNG}}
\efinew

However, differences in the cosmological parameters of these runs are small. To test the generality of the fitting formulas against cosmological parameter variations, we investigate a cosmology variation series.
\begin{itemize}
    \item The cosmology variation series (8 runs). We choose the 1P set of the CAMELS-TNG runs identifiers 1P\_pX\_Y (X=$1,2,7=\Omega_m,\sigma_8, \Omega_b$. Y$=2$, n2).  We also include 1P\_p1\_1 and 1P\_p1\_n1. $\Omega_m=0.1,0.2,0.4,0.5$. $\sigma_8=0.6,1.0$ and $\Omega_b=0.029,0.069$. Note that the fiducial cosmology run with $\Omega_m=0.3,\sigma_8=0.8$ and $\Omega_b=0.049$ is already included in the TNG series previously described. 
\end{itemize}

\section{Road to the single-parameter fitting formula}
\label{sec:1P}

In the beginning of this work, we tested against the FLAMINGO simulation suite and made the choice of 
\begin{equation}
    n=1\ .
\end{equation}
We will stick to this choice in this section, and construct the single-parameter formula by hierarchical fitting. This key simplification will be relaxed in the three-parameter form. Furthermore, originally we choose to fit at $k\leq 3h/$Mpc, since this is the range that weak lensing cosmology concerns the most. Unless otherwise specified, we stick to this original choice.\footnote{Nonetheless, we will show that the fitting is applicable to $k\la 5h/$Mpc, with essentially the same accuracy (Fig. \ref{fig:3Pk5k1345}).  But beyond $\sim 5h/$Mpc gas cooling induced power spectrum enhancement starts to dominate over feedback induced suppression. Our fitting formula monotonically decreases with $k$. So this turnover scale at $k\sim 5h/$Mpc sets the upper $k$ limit of our fitting formulas. }

\bfinew[width=1.0\textwidth]{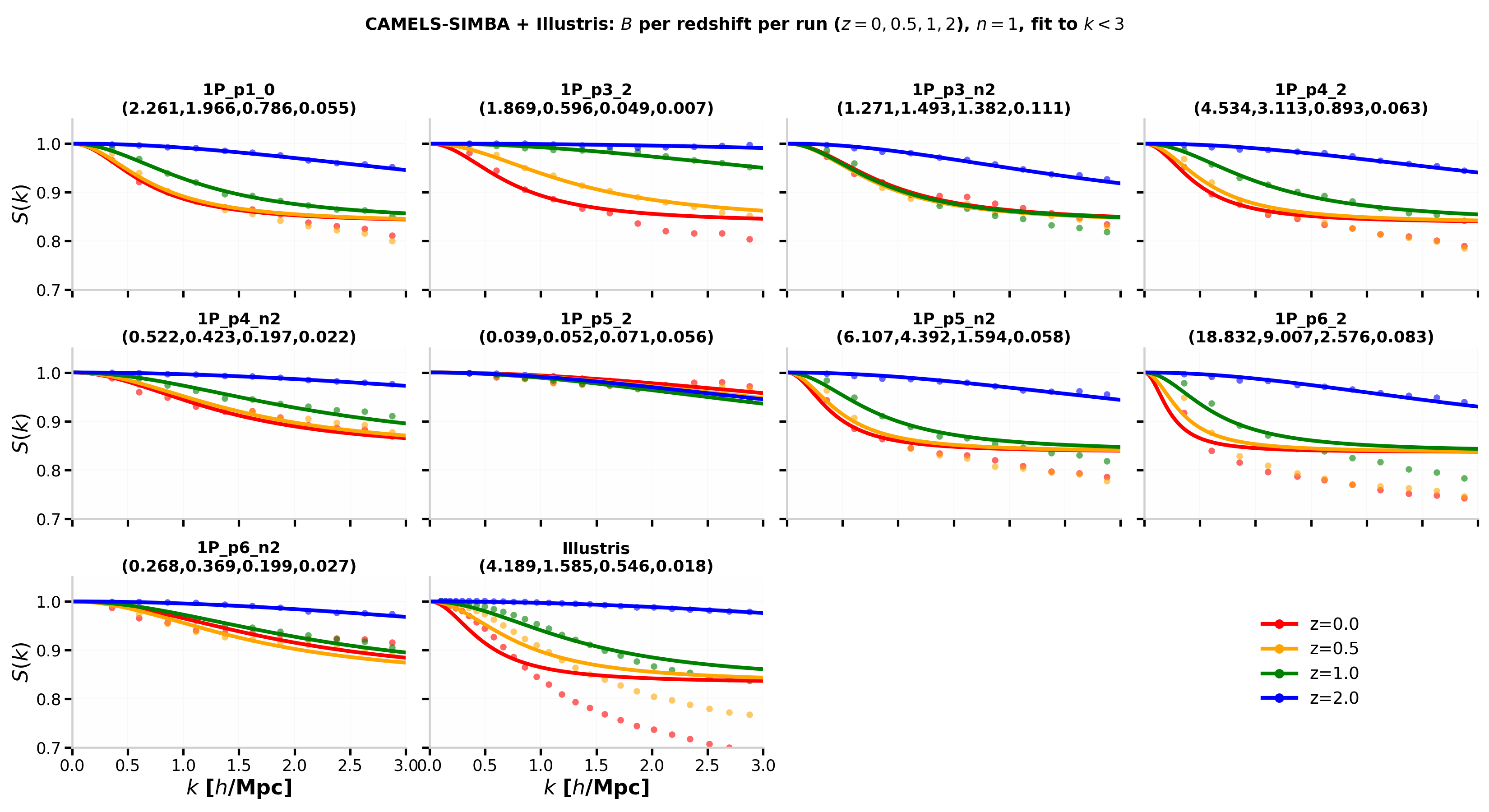}
\caption{The same as Fig. \ref{fig:FLAMINGO} \& \ref{fig:TNG}, but for the strong feedback series. The most dramatic feedback cases such as Illustris and CAMELS-SIMBA 1P\_p6\_2 clearly reveal two limitations of Eq. \ref{eqn:Sfit} with $n=1$. First is an overly high suppression floor of $f_{\rm dm}=1-\Omega_b/\Omega_m\sim 0.83$. Second is the mismatched shape at $k<1h/$Mpc. These behaviors imply the existence of new effective DoFs. The fitting formula taking them into account describes these simulation runs excellently (Fig. \ref{fig:varyingn}). \label{fig:extreme} }
\efinew

\subsection{Parameterization of $B$}
First we fit $B(z_i)$ per redshift per run ($z_i=0,0.5,1,2$) using Eq. \ref{eqn:Sfit}. The best-fit $B$ varies significantly across redshifts and simulations (Fig. \ref{fig:FLAMINGO}, \ref{fig:TNG} \& \ref{fig:extreme}). The maximum value of  $B$ is $\sim 10^{-1}$ for the TNG series (Fig. \ref{fig:TNG}). It increases to $\sim 10^{-1}$-$10^0$ for the FLAMINGO series (Fig. \ref{fig:FLAMINGO}). Illustris and many of CAMELS-SIMBA series have $B_{\rm max}\sim 1$-$10$ (Fig. \ref{fig:extreme}). 

It is more instructive to check in terms of $l_B\equiv B^{1/2}$, which denotes the typical gas particle displacement scale. The typical value of the maximum $l_B$ in the TNG series is $\sim 0.3 $ Mpc$/h$. This size of displacement of gas particles  is comparable to  the virial radius of $10^{13}M_\odot$ halos and therefore would have significant impact on their gas fraction. For the extreme feedback series, the maximum $l_B$ occurs today, with an amplitude of $\sim 1$ Mpc$/h$. This could even significantly reduce the gas fraction in $10^{14.5}M_\odot$ halos.\footnote{Note that some runs in Fig. \ref{fig:extreme} have unrealistically large $l_B$ (e.g. $\sim 4$ Mpc$/h$), due to the inappropriate choice of $n=1$. By relaxing $n$, the maximum $l_B$ reduces to $1$-$2$ Mpc$/h$ (Fig. \ref{fig:varyingn}), corresponding to the virial size of $10^{14}$-$10^{15}h^{-1}M_\odot$ halos. }

Despite the above diversity in the value of $B$, its redshift evolution is well described by Eq. \ref{eqn:B}, namely
\begin{equation}
\label{eqn:BzB0}
    B(z)=B_0(1+z)^ae^{-bz}\ \Leftrightarrow\ l_B(z)=l_0 (1+z)^{a/2}e^{-bz/2}\ .
\end{equation} 
This redshift dependence reflects the competition between energy and momentum injection which displace gas particles and gravitational force which pulls them back. 

$b$ varies across simulations noticeably. However, we find that approximating $b$ as a global parameter $\sim 4$ is acceptable in fitting $S(k,z)$. 
We then fit against $B(z_i)$ of each run for the two free parameters $(B_0,a)$, together with a global parameter $b$.\footnote{Note that we do not fit the cosmology variation runs here. Instead we use the best-fit $b$ to test its generality against the cosmology variation runs.} Fig. \ref{fig:Bz} shows that the functional form of Eq. \ref{eqn:B} is excellent. The best-fit $b$ is
\begin{equation}
    b=3.87\ .
\end{equation}
 We notice a possible connection between the value of $b$ and  the growth rate of $10^{13} h^{-1}M_\odot$ halos. We define $\alpha\equiv -d\ln n_h(>M)/dz$. The growth of halo mass function $n_h(>M)$ is then approximately $\propto \exp(-\alpha z)$. Using the CSST halo mass function emulator \cite{2025SCPMA..6809513C}, we find that $\alpha\sim 2\sim b/2$ for $M\sim 10^{13} h^{-1}M_\odot$ at $z\sim 1$. 
The term $\exp(-bz/2)$ in $l_B(z)$ (Eq. \ref{eqn:BzB0})  describes the growth of the gas particle displacement scale $l_B$. If AGN feedback dominates in this process and if the AGN feedback  is dominated by $10^{13}h^{-1}M_\odot$ halos, this could be a plausible explanation of $b\sim 4$.  Nevertheless,  it is unclear whether $b$ is related to the above process and it is beyond the scope of this paper to interpret $b$. 

In contrast, $B_0$ varies by more than three orders of magnitude, from $\sim 10^{-2}$ to $\sim 10$ ($({\rm Mpc}/h)^2$). Therefore we identify $B_0$ as a key DoF in describing $S(k)$.

\bfinew[width=1.0\textwidth]{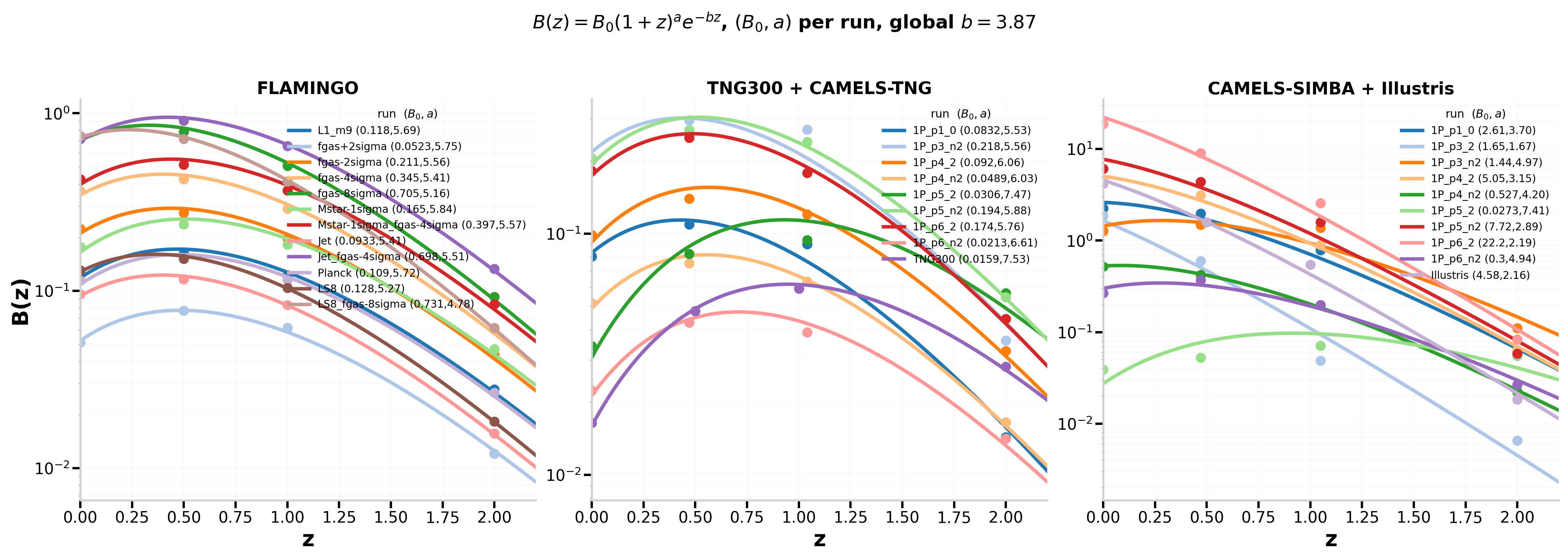}
\caption{$B(z)$ is well fitted by $B(z)=B_0(1+z)^a\exp(-bz)$ (Eq. \ref{eqn:B}) with a global $b=3.87$.  \label{fig:Bz}}
\efinew

\bfinew[width=0.49\textwidth]{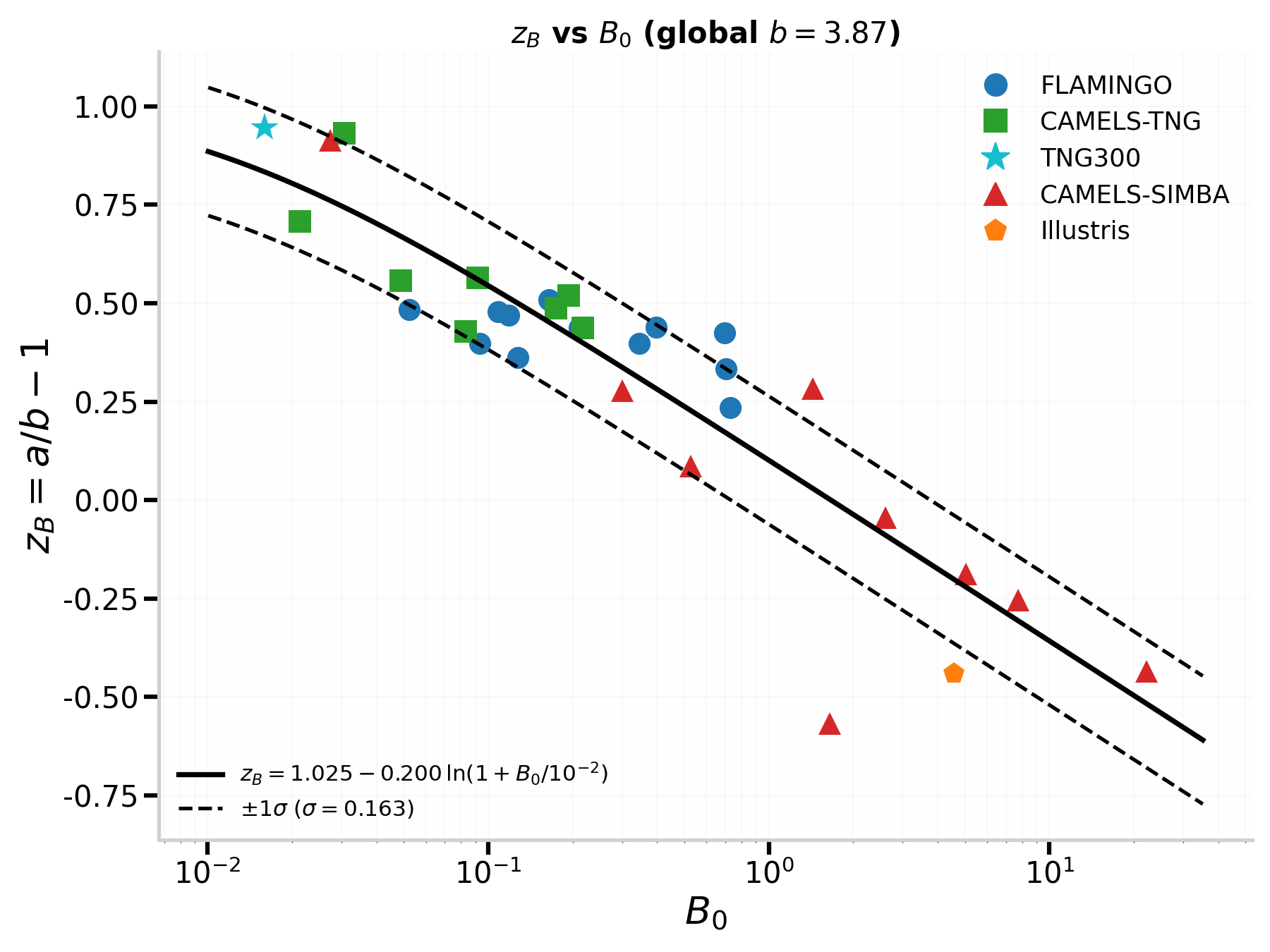}
\includegraphics[width=0.49\textwidth]{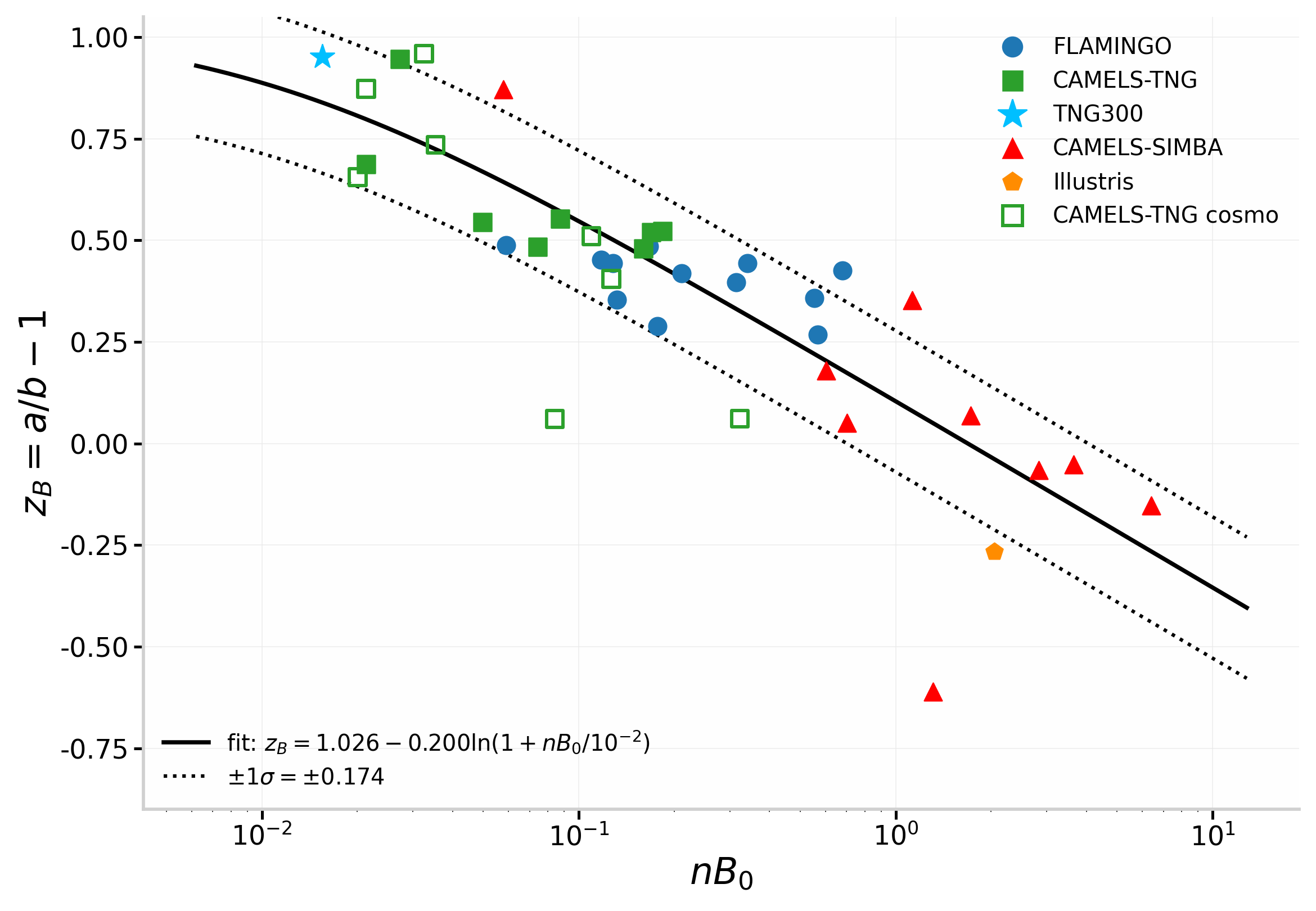}
\caption{Left panel, the $B_0$-$z_B$ scatter plot from the two-parameter ($B_0$, $a$) fitting of 32 runs (not including the CAMELS-TNG cosmology variation) shown in Fig. \ref{fig:FLAMINGO}, \ref{fig:TNG} \& \ref{fig:extreme}. $z_B=a/b-1$ is the redshift of peak feedback suppression.  An interesting finding is that  the strength and the history of the suppression ($B_0$ and $z_B$) are correlated. This is a key ingredient in constructing the single-parameter fitting formula. However, scatters and outliers of this relation (e.g., CAMELS-SIMBA 1P\_p3\_2), indicate that a more accurate fitting formula should include $z_B$ (or equivalently $a$) as an effective DoF.   Right panel, the $nB_0$-$z_B$ scatter plot from the three-parameter ($B_0$,$a$,$n$) fitting. The best-fit relation is essentially identical, by replacing $B_0$ in the $n=1$ case (left panel) with $nB_0$. The correlation is still significant, but not sufficiently tight to fix $z_B$.  We also show the data points of CAMELS-TNG cosmology variation runs not used in the fitting.  \label{fig:zB}}
\efinew

\subsection{The peak redshift-suppression strength relation}
The variation of $a$ falls between that of $b$ and $B_0$. $a$ determines  the redshift $z_B$ where $B(z)$ (or equivalently $l_B(z)$) reaches maximum. 
\begin{equation}
    z_B=\frac{a}{b}-1\ .
\end{equation}

Within the FLAMINGO series,  $z_B\sim 0.3$ and it does not vary dramatically. Within the CAMELS-TNG series, $z_B\sim 0.5$. For both series, the peak suppression occurred in the past. However, for Illustris, $B$  keeps increasing with time and is expected to peak in the future. This is also the case for some of the CAMELS-SIMBA runs.  Therefore $z_B$ (or equivalently $a$) contains key information on the feedback history. 

Fig. \ref{fig:zB} shows that $z_B$  is correlated to  $B_0$. The stronger the feedback, the later the peak $z_B$. To avoid fitting instabilities caused by poorly determined $B_0$ in the case of weak suppression,  we choose to fit  the $B_0$-$z_B$ relation with the following format,
\begin{equation}
\label{eqn:zBB0}
    z_B=c_0+c_1\ln (1+B_0/10^{-2})\ .
\end{equation}
The best-fit values are
\begin{equation}
    c_0=1.03\ ,\ c_1=-0.20\ .
\end{equation}
The correlation is significant, and persists when we relax $n=1$. 

Exactly what causes  is an interesting topic of further study. 

However, we note that there are significant scatters and outliers to the above relation ($\sigma(z_B)\sim 0.16$). Relevant outliers beyond 1-$\sigma$ (left panel, Fig.\ref{fig:zB}) are strong feedback runs including Illustris and two CAMELS-SIMBA runs, with the CAMELS-SIMBA 1P\_p3\_2 as the most significant (more than 3-$\sigma$). What causes the above correlation and what causes these outliers are interesting issues for further study.  

\subsection{Performance of the single-parameter fitting formula}
Then finally we arrive at the single-parameter fitting formula (Eq. \ref{eqn:Sfit}, \ref{eqn:B} \& \ref{eqn:1P}), in which $B_0$ is the only free parameter. Note that the parameters ($b$, $c_0$, $c_1$) are determined by hierarchical fittings described previously, instead of directly fitting $S(k,z)$.  For a global fitting, the best-fit $b$, $c_0$ and $c_1$ indeed change noticeably. However, in terms of the $S(k)$ fitting accuracy, the impact is negligible. Therefore we will stick to their values determined by the above hierarchical fitting procedure, which is more instructive. 

With dramatically reduced degrees of freedom (DoFs), the fitting accuracy of this single-parameter recipe degrades compared to that in Fig. \ref{fig:FLAMINGO}, \ref{fig:TNG} \& \ref{fig:extreme}.  Nonetheless, the fitting accuracy is still impressive (Fig. \ref{fig:DS1} \& \ref{fig:DS2}). We further check its cosmological generality by fitting the cosmology variation series (Fig. \ref{fig:TNGcos}). These simulation runs have significantly different cosmological parameters from the training set (e.g. $\Omega_m=0.1$, $\sigma_8=0.6$ or $\Omega_b=0.069$). The single-parameter fitting formula works surprisingly well and $f_{\rm dm}$ absorbs almost all cosmological dependence. The worst performance occurs for  the 1P\_p1\_n2 run, due to  unrealistically high baryon fraction $\Omega_b/\Omega_m=0.49$. Nonetheless, the fitting formula works reasonably well. 

\bfinew[width=1.0\textwidth]{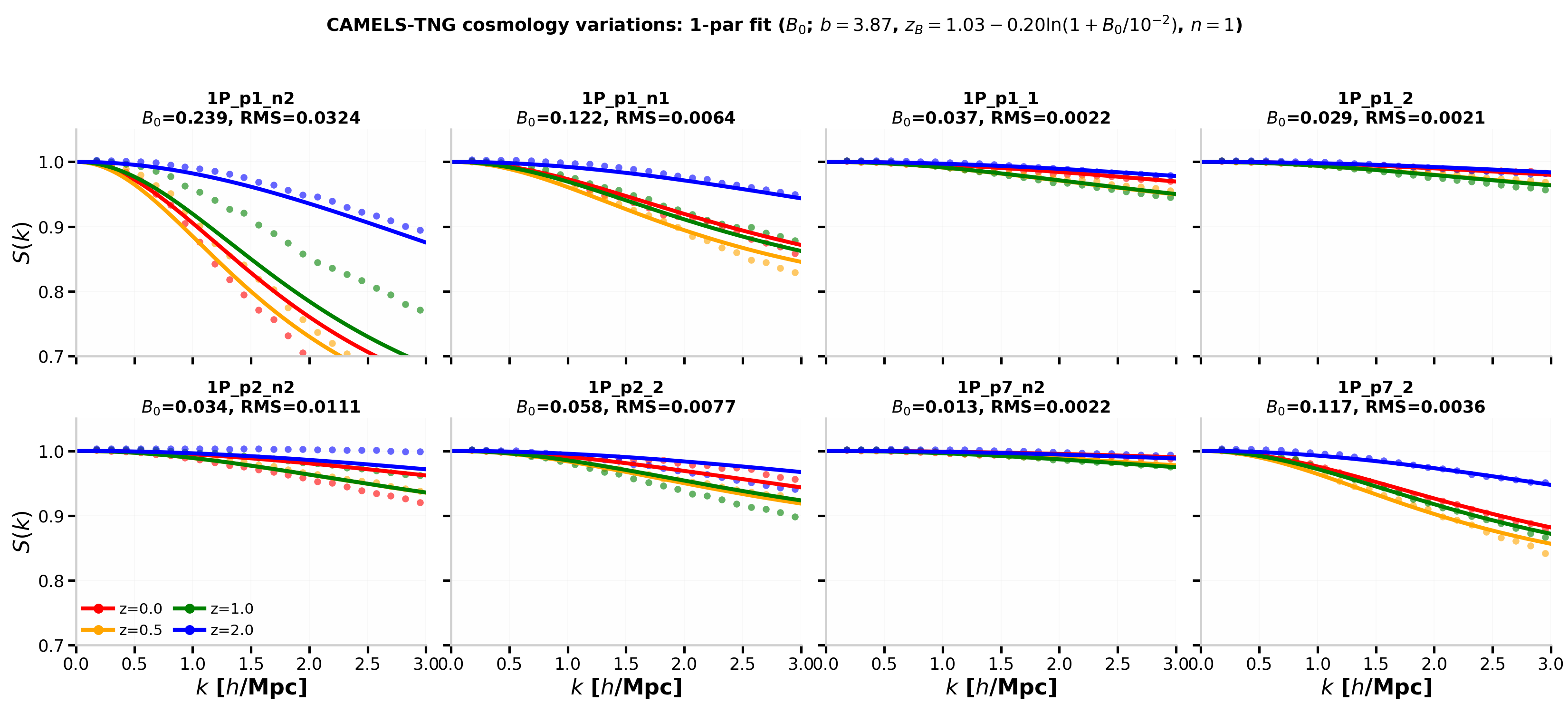}
\caption{Performance of the single-parameter fitting formula against the CAMELS-TNG 1P set of cosmology variations. Large error occurs only for unrealistically low $\Omega_m=0.1$ (and therefore unrealistically high baryon fraction of $49\%$). Relaxing $n=1$ and relaxing the $B_0$-$z_B$ relation further improve the performance.   \label{fig:TNGcos}}
\efinew

\bfinew[width=1.0\textwidth]{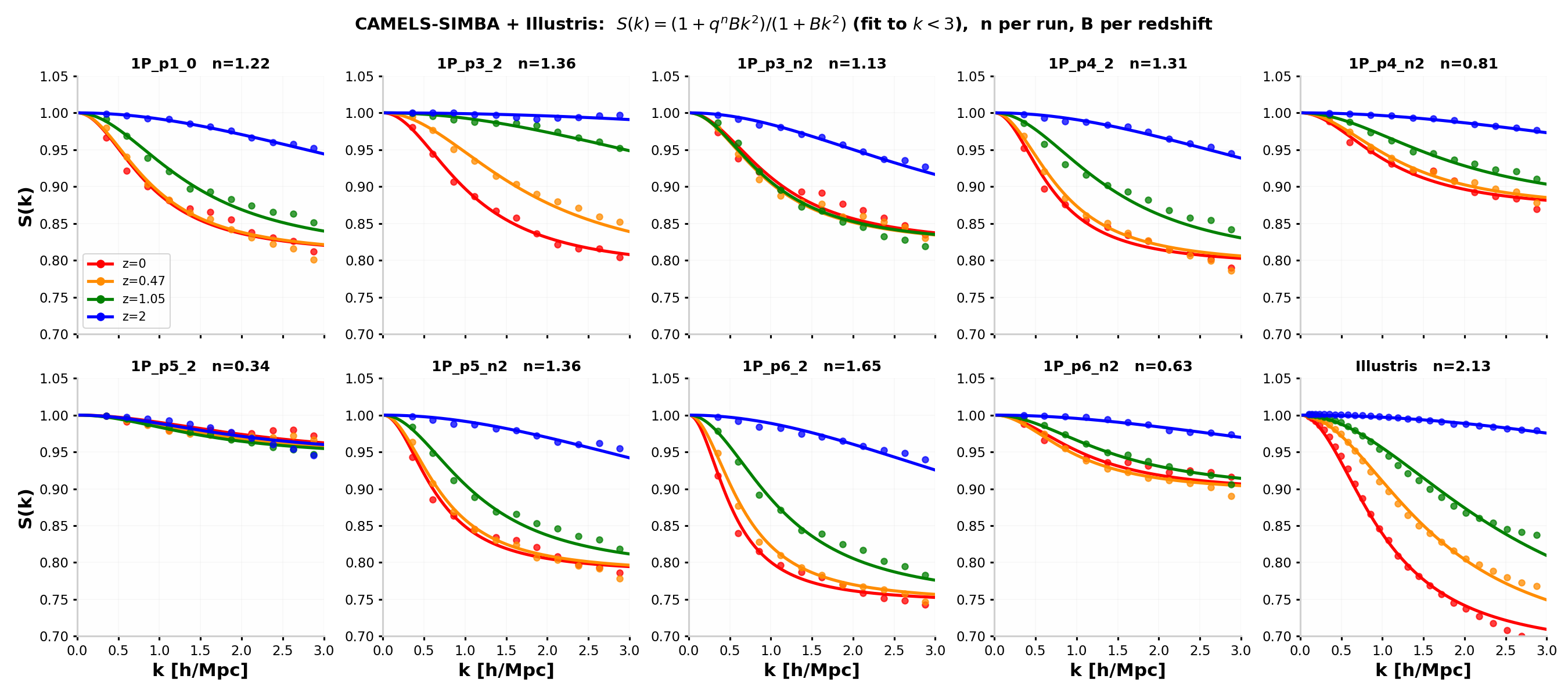}
\caption{Relaxing $n=1$ and the $B_0$-$z_B$ relation significantly improve the fitting accuracy.  Compared to Fig. \ref{fig:extreme}, we need to relax $n=1$ in order to match the low $S\sim 0.7$ behavior of Illustris and CAMELS-SIMBA 1P\_p6\_2. Compared to Fig. \ref{fig:zB}, we need to relax the $B_0$-$z_B$ relation to match the redshift evolution. The performance for other runs such as FLAMINGO and TNG is better. Therefore we only show the performance of the strong feedback runs.    \label{fig:varyingn}}
\efinew

\section{The three-parameter fitting formula}
\label{sec:3P}
Fig. \ref{fig:extreme} \& \ref{fig:zB} reveal two extra DoFs essential to improve the fitting formula accuracy, in particular at $k>1h/$Mpc and/or for cases of strong feedback. 

A fundamental drawback of fixing $n=1$ in Eq. \ref{eqn:Sfit} is the suppression floor $S\geq f_{\rm dm}=1-\Omega_b/\Omega_m>0.83$. Therefore it fails to describe the deep suppression $S\sim 0.7$ of Illustris and some CAMELS-SIMBA runs (e.g. 1P\_p6\_2). By varying $n$, this suppression floor issue can be naturally resolved. For $n=2$, the floor is now $S=f_{\rm dm}^2\sim 1-2(\Omega_b/\Omega_m)\sim 0.7$, meeting the requirements from Illustris and CAMELS-SIMBA.  

Relaxing $n$ from $n=1$ also changes the best-fit of $B$ at each redshift. The two together make  Eq. \ref{eqn:Sfit} versatile in describing both the shape and amplitude of the simulated $S(k)$. Fig. \ref{fig:varyingn} shows that the description is accurate to $k=3h/$Mpc, for Illustris and CAMELS-SIMBA variations. Interestingly, for Illustris, the best-fit $n=2.13$ instead of $n=2$. This is likely caused by the extra suppression in the density growth due to the back-reaction of smoothed baryon distribution. The performance for other simulations is even better. We conclude that the parameter $n$ and the function $B(z)$ contain all effective DoFs in the shape ($k$ dependence) of $S(k,z)$. 

$B(z)$ is not a completely free function of $z$. We have shown that it is well described by $B(z)=B_0(1+z)^a\exp(-bz)$ with $b=3.87$. The $B_0$-$z_B$ relation (Fig. \ref{fig:zB}) further compresses the whole redshift evolution into a single parameter $B_0$. However, the CAMELS-SIMBA 1P\_p3\_2 outlier point in the $B_0$-$z_B$ relation (Fig. \ref{fig:zB}) shows that this relation is not sufficiently accurate. Actually, it is responsible for the significant underestimation of $S$ in Fig. \ref{fig:DS2}.  

Furthermore, we find no significant correlation  between scatters in the $B_0$-$z_B$ plot and $n$. Then to improve the fitting accuracy, we have to relax the $B_0$-$z_B$ relation (Eq. \ref{eqn:zBB0}) and treat $B_0$ and $z_B$ (or equivalently $a=b(1+z_B)$) as two independent free parameters. 

Finally we arrive at the three-parameter fitting formula, with $(B_0,a,n)$ as the three free parameters. $n$ controls how deep the suppression can go. $B_0$ controls the transition scale of the suppression function, which sets the feedback strength. $n$ and $B_0$ together account for all the effective DoFs in the shape of $S(k)$. $a$ (or equivalently $z_B$) describes the history of feedback and accounts for the effective DoF in the redshift evolution. 

This three-parameter fitting formula works excellently for all investigated feedback scenarios, from moderate ones such as TNG and FLAMINGO to strong ones such as Illustris. If we treat all simulation data points with equal weight, the mean bias in $S$ is negligible ($\sim 10^{-3}$) and the r.m.s. across runs is well below $10^{-2}$ at $k=3h/$Mpc (Fig. \ref{fig:3Perrork3}). A more meaningful measure is the maximum error in $S$. For the 32 runs to  calibrate the fitting formula, max$(|\Delta S|)<0.03$ at $k=3h/$Mpc. The formula performs equally well for the CAMELS-TNG cosmology variations in the range $\Omega_m\in [0.2,0.5]$, $\Omega_b\in [0.029,0.069]$ and $\sigma_8\in [0.6,1.0]$. Nonetheless, when applied to the case of $\Omega_m=0.1$, max$(|\Delta S|)\simeq 0.05$ occurs. However, this relatively large error is essentially irrelevant for cosmology. First, the value of $\Omega_m=0.1$ ($\Omega_b/\Omega_m=0.49$) is unrealistically low(high). So it should be solely regarded as a stress test. Second, this large error occurs at $z=2$. It is irrelevant for the majority of weak lensing measurements, whose source redshift is well below 2. Even for source redshift beyond $z=2$, its impact is heavily suppressed by the weak lensing kernel. In the redshift range $z<1$ of most interest to weak lensing cosmology, $|\Delta S|\leq 0.02$ even for $\Omega_m=0.1$.

There also exists a correlation between $nB_0$ and $z_B$ and the best fit relation is essentially identical (Fig. \ref{fig:zB}), so we can adopt Eq. \ref{eqn:zBB0} and simply replace $B_0$ with $nB_0$.
\begin{equation}
\label{eqn:zBnB0}
    z_B=c_0+c_1\ln(1+nB_0/10^{-2})\ ,\ c_0=1.03\ ,\ c_1=-0.20\ .
\end{equation}
Adopting this relation, the three-parameter formula reduces to a two-parameter formula. Unfortunately, this two-parameter formula fails to compete with the three-parameter one for accuracy and applicability, due to scatters and outliers in the $nB_0$-$z_B$ relation (Fig. \ref{fig:zB}). It also fails to compete with the single-parameter formula for the simplicity. Its performance falls between the three-parameter and single-parameter formulas, and implies that two effective DoFs are not sufficient to accurately describe $S(k,z)$. Nevertheless, the existence of the $nB_0$-$z_B$ relation implies that the number of effective DoFs required for accurate description of $S(k,z)$ is somewhere between two and three. 

\section{Discussions}
\label{sec:discussion}
We have constructed a three-parameter ($B_0$, $n$ and $a$) fitting formula to describe the baryonic feedback suppression $S(k,z)$ of the matter power spectrum. $B_0$  quantifies the typical gas particle displacement $l_B\equiv B_0^{1/2}$, $n$ sets the suppression floor $f_{\rm dm}^n=(1-\Omega_b/\Omega_m)^n$, and $a$  quantifies the redshift $z_B$  at which the suppression peaks. We also find a relation between $z_B$ and $B_0$. Combined with the choice of $n=1$, we obtain a single-parameter reduction. Their performance is summarized as follows.
\begin{itemize}
    \item The three-parameter formula. The maximum error at $k\leq 3h/$Mpc and $z\leq 2$ is below $0.01$ for weak-to-moderate feedback scenarios such as TNG and FLAMINGO (both thermal and jet AGN feedback modes), and below $0.03$  for strong feedback scenarios such as Illustris and some CAMELS-SIMBA runs.
    \item The single-parameter formula. It is accurate for TNG and FLAMINGO, with max$(|\Delta S|)<0.01$ at $k=1h/$Mpc and max$(|\Delta S|)<0.03$ at $k=3h/$Mpc. It is applicable to strong feedback scenarios at $k\leq 1h/$Mpc, with max$(|\Delta S|)<0.04$. 
    \item Both formulas are robust to cosmology variations, and the $f_{\rm dm}^n$ term  absorbs essentially all cosmological dependence. Within the tested CAMELS-TNG cosmology variation runs, the only exception is the run with  unrealistically low $\Omega_m=0.1$. Even for this extreme case, the two formulas still work at $k\leq 1h/$Mpc.      
\end{itemize} 
Therefore we are able to conclude that there are at most three effective DoFs in $S(k,z)$. This finding is broadly consistent with several findings from independent viewpoints. Principal component analyses of baryonic-physics-induced changes to the weak lensing angular power spectrum found that marginalization over at most the first 3-4 (or even 2) PCs is sufficient for Stage IV weak lensing surveys such as LSST \cite{2015MNRAS.454.2451E,2025JCAP...03..041R}. In weak-lensing-only analyses of Stage III surveys based on halo-model methods, only one baryonic physics parameter (the halo mass at which half of the cosmic gas fraction is expelled) is enough (e.g., DES-Y3, KiDS-1000 and HSC-DR1, \cite{2024JCAP...08..024G}). However, the one-parameter recipes such as that of BCEmu cannot reach $1\%$ accuracy in $S(k)$ and the three-parameter versions are required for the joint analysis of weak lensing and kSZ \cite{2024MNRAS.534..655B}.
Recently, a machine-learning analysis found a unified 2D latent representation of baryonic physics, independent of both time and cosmology \cite{2026ApJ...996L..41L}.

It is instructive to compare with existing modeling of $S(k)$, which can be classified into several categories. 
\begin{itemize}
    \item Analytical fitting formulas requiring no external input. Our model belongs to this category. An existing fitting formula in this category is the $A_{\rm mod}(k,z)$ model proposed by \cite{2022MNRAS.516.5355A} and then parameterized and calibrated against the FLAMINGO suite \cite{2025MNRAS.540.2322S}. Interestingly, it also has a one-parameter version and a three-parameter version \cite{2025MNRAS.540.2322S}. The one-parameter version has limited applicability (e.g., limited to $z<1$ and unable to reproduce the jet-like AGN feedback variants). The three-parameter version is applicable to all FLAMINGO runs, however its applicability at $z>1$ and to simulations beyond FLAMINGO requires further quantification. In comparison, our single-parameter formula applies to $z\leq 2$  and to both thermal and jet AGN feedback modes of FLAMINGO runs, and our three-parameter formula applies to simulations beyond FLAMINGO (e.g., Illustris). 
    \item Analytical fitting formulas requiring external  input. \cite{2020MNRAS.491.2424V} found a quite generic relation between the baryon fraction of $\sim 10^{14}M_\odot$ galaxy clusters  and  the feedback suppression strength at $k\sim 1h/$Mpc, insensitive to detailed baryonic physics. This finding motivated fitting formulas to map the observed baryon fraction of galaxy groups and clusters to $S(k)$, including the SP(k) model \cite{2023MNRAS.523.2247S} and the resummation model \cite{2024MNRAS.528.4623V,2026MNRAS.545f2086V}. The mapping is calibrated against hydro simulations in the hope that the mapping is generic.   The mapping between the baryon fraction and the power spectrum response may involve many parameters. But once calibrated against chosen simulations, these models in principle contain no free parameters and are expected to be applicable to baryonic physics scenarios beyond the calibration simulations. Nevertheless, their robustness to variations in cosmology, in particular $\Omega_b/\Omega_m$,  beyond the range probed by FLAMINGO  remains to be tested \cite{2026MNRAS.545f2086V}. This cosmological dependence is absorbed explicitly by our formula  through the $f_{\rm dm}^n$ term.
Furthermore, a major challenge when applying these models is the accurate measurement of baryon fractions across halo mass and redshift. On one hand, we have multiple observable proxies, such as tSZ, kSZ and X-ray  (e.g., \cite{2022MNRAS.514.3802S,2023ApJ...953..188C,2024MNRAS.528.4379G,2024MNRAS.534..655B,2025PhRvD.112h3509H,2026MNRAS.550g1314B,2026JCAP...03..036D,2026MNRAS.549ag993S}). On the other hand, converting these proxies into gas fraction and then baryon fraction is in general non-trivial. Alternatively, there are proposals to connect observable proxies directly to $S(k)$. For example, \cite{2026PhRvD.113h3507L} trained a neural network using FLAMINGO to translate kSZ into $S(k)$. Furthermore, the rapidly growing sample of fast radio bursts (FRBs; \cite{2023RvMP...95c5005Z}) is enabling large scale structure analyses \cite{2026arXiv260422105S,2026arXiv260704106W}. Eventually, localized FRBs will enable a model-independent measurement of $S$ and the reconstruction of a dark-matter-only universe at field level \cite{2025arXiv251110975Z}. 
    \item Emulators. Both BCEmu \cite{2021JCAP...12..046G,2023ascl.soft08010G} and BACCO \cite{2021MNRAS.506.4070A} are emulators for $S(k)$ trained on baryonified simulations. The default versions therefore have 7 baryonic physics parameters at each redshift. Both provide reduced-parameter versions. For example, the 3-parameter version of BACCO achieves better than $3\%$($5\%$) accuracy at $k=1h/$Mpc and $z=0$($z=1$), for the simulations tested (Fig. 7 \& 8 in \cite{2021MNRAS.506.4070A}). The 7-parameter version of BACCO improves slightly. Furthermore, these versions tend to underestimate $S(k)$ at $k\sim 0.5h/$Mpc by a few percent. BCEmu has a 4-parameter version including a parameter quantifying the redshift dependence. The performance (Fig. 11, \cite{2021JCAP...12..046G}) is comparable to our three-parameter formula.   Nonetheless, since BACCO and BCEmu  were tested against a different set of simulations, we do not directly compare them with ours.    Recently, \cite{2025MNRAS.539.1337S} constructed an emulator of $S(k)$ ($R(k)$ in their notation) calibrated with FLAMINGO. This emulator has three free parameters of cluster gas fraction, jet fraction and stellar mass. It is accurate to better than 1\%  up to $k=10h$/Mpc, and is therefore able to describe the small scale power spectrum enhancement that our formula cannot (and is not intended to) describe. It shows applicability to baryonic physics scenarios beyond FLAMINGO, but exceptions (e.g., Illustris in their Fig. 11) also exist.  As a comparison, our three-parameter formula is applicable to a wider range of cosmologies than the FLAMINGO-calibrated emulator, and likely to a wider range of baryonic physics scenarios.
    \item Semi-analytical models. They include HMCODE \cite{2015MNRAS.454.1958M,2021MNRAS.502.1401M} and baryonification \cite{2015JCAP...12..049S,2021MNRAS.503.3596A,2025JCAP...12..043S,2025JCAP...11..046K}, both of which are halo-model-motivated.  The baryonic feedback model in HMCODE-2020 contains 6 free parameters characterizing the gas fraction, stellar mass fraction, change of halo concentration, and their redshift dependence. HMCODE-2020 also has a single-parameter version, with $T_{\rm AGN}$ as the only free parameter. This version was calibrated on BAHAMAS alone, whereas our single-parameter formula is validated on multiple suites including jet-mode AGN feedback and to $z<2$. The updated version of baryonification \cite{2025JCAP...11..046K} has 8 parameters. Its 2+1 parameter version achieves better than $2\%$ accuracy at $k\leq 3h/$Mpc for FLAMINGO and TNG. This is comparable to our 3-parameter formula. However, it has not been compared against simulations with stronger feedback or significantly different cosmologies yet. We note that, unlike our formula, the halo-model-based frameworks can be extended to summary statistics beyond $S(k)$, such as the DMO power spectrum, the SZ power spectrum,  and even to field-level information in the case of baryonification. 
\end{itemize}
We note that the comparisons above are indicative only, since the performance of these models has not been tested against the same simulations at the same $k$ and $z$ range. Nevertheless, we demonstrate that our fitting formulas are competitive with existing ones. Furthermore, these efforts  will not only provide a cross-check of this key systematic uncertainty in weak lensing cosmology, but also yield useful information on baryonic physics. 

Finally,  we caution that in our analysis, we neglect statistical uncertainties and systematic errors in the simulated power spectra and $S(k)$ (e.g., the boxsize-induced systematic errors, \cite{2025MNRAS.539.1337S}). We fit the data at face value and with equal weight. Therefore the performance of the fitting formula and also the prefixed parameters (e.g., $b=3.87$) should be re-calibrated as simulation data are updated and improved. Another issue is that the simulation test sample is still limited and future work should enlarge the simulation sample to further test the generality of the proposed fitting formula.  Furthermore, the impact of an error in $S$  at a given $k$ and $z$  on cosmology depends on the survey specifications and the cosmological parameters of interest. The overall impact on cosmology should  be weaker than that indicated by the quoted max$(|\Delta S|)$, which only applies to a small fraction of data points. We leave detailed investigation  for specific surveys and cosmological parameters  to future work. 

{\bf Acknowledgment}.--- This work was supported by NSFC (12595310) and the National Key R\&D Program of China (2023YFA1607800, 2023YFA1607801). Numerical calculations were performed with the web version of the Kimi K3 Large Language Model (\url{https://www.kimi.com/}). We acknowledge the Virgo Consortium for making their simulation data available. The FLAMINGO simulations were performed using the Durham Memory Intensive system managed by the Institute for Computational Cosmology on behalf of the STFC DiRAC facility (\url{www.dirac.ac.uk}).

\bibliography{feedbacksuppression} 

\appendix
\section{The best-fit values of the three-parameter fitting}
\label{sec:n}
For the three-parameter formula, we list the best-fit values of the three parameters $B_0$, $a$ and $n$ for each of the 40 simulation runs (Table \ref{tab:3P}). As a reminder, $S(k)=(1+f_{\rm dm}^n Bk^2)/(1+Bk^2)$, $B(z)=B_0(1+z)^a e^{-bz}$, and $b=3.87$ is fixed.  The fitting uses all the data points at $k\leq 3\,h/{\rm Mpc}$ with equal weight, for 4 redshifts of  each run. The fitting is excellent. The r.m.s. fitting error of $S(k,z)$ is usually less than  $3\times 10^{-3}$ for each run in the FLAMINGO and TNG series. It increases for Illustris and CAMELS-SIMBA, which is usually smaller than $10^{-2}$. The maximum r.m.s. error ($0.0126$) occurs for the CAMELS-TNG cosmology run with $\Omega_m=0.1$. 

A finding is that $n$ varies with feedback scenarios. For weak-to-moderate feedback scenarios, $n\sim 1$. Stronger feedback prefers $n>1$. For example,  Illustris has $n\simeq 2$. At the same time, there is  visible scatter in $n$. This indicates that $n$ captures an effective DoF of the feedback. Another finding is that the peak suppression redshift $z_B\equiv a/b-1$ is correlated with $nB_0$ (right panel, Fig. \ref{fig:zB}). The relation is almost identical to that in the case of $n=1$, despite significant changes in the best-fit $B_0$. 

\begin{table}
\centering
\caption{Best-fit parameters of the three-parameter fit. The r.m.s. per run is in general much smaller than $10^{-2}$ and only two runs have r.m.s. greater than $0.01$ ($0.0116$ for the CAMELS-SIMBA run 1P\_p3\_n2 and $0.0126$ for the CAMELS-TNG 1P\_p1\_n2 run with $\Omega_m=0.1$). Note that for $B_0\la 10^{-2}$, there is a strong degeneracy of the form $(1-f_{\rm dm}^n)B\sim nB(\Omega_b/\Omega_m)$ and two of the runs hit the boundary of the adopted prior $n\in[-3,4]$. The one having $n=-3$ actually has power enhancement instead of power suppression. The quoted $B_0$ and $n$ for these two runs should be used with caution. Nonetheless, their values have little impact on the fitting accuracy of $S(k)$, since $S(k)\simeq 1$ due to small $B_0$. Also note that $n=2.11$ of Illustris differs from $n=2.13$ quoted in \S \ref{sec:3P}. The one in \S \ref{sec:3P} fits $B(z)$ per redshift, which is different from the fitting with a functional form of $B(z)$ presented here. \label{tab:3P} }
\begin{tabular}{llcccc}
\hline\hline
Suite & Run & $B_0$ & $a$ & $n$ & r.m.s. \\ \hline
FLAMINGO & L1\_m9 & 0.145 & 5.59 & 0.89 & 0.0021 \\
FLAMINGO & fgas+2sigma & 0.088 & 5.76 & 0.67 & 0.0021 \\
FLAMINGO & fgas-2sigma & 0.206 & 5.49 & 1.02 & 0.0022 \\
FLAMINGO & fgas-4sigma & 0.279 & 5.40 & 1.13 & 0.0023 \\
FLAMINGO & fgas-8sigma & 0.448 & 5.25 & 1.24 & 0.0024 \\
FLAMINGO & Mstar-1sigma & 0.164 & 5.75 & 1.01 & 0.0024 \\
FLAMINGO & Mstar-1sigma\_fgas-4sigma & 0.285 & 5.59 & 1.19 & 0.0026 \\
FLAMINGO & Jet & 0.377 & 4.99 & 0.47 & 0.0026 \\
FLAMINGO & Jet\_fgas-4sigma & 0.667 & 5.52 & 1.02 & 0.0029 \\
FLAMINGO & Planck & 0.133 & 5.62 & 0.88 & 0.0021 \\
FLAMINGO & LS8 & 0.140 & 5.24 & 0.94 & 0.0022 \\
FLAMINGO & LS8\_fgas-8sigma & 0.449 & 4.90 & 1.26 & 0.0021 \\
\hline
CAMELS-TNG & 1P\_p1\_0 & 0.060 & 5.74 & 1.24 & 0.0025 \\
CAMELS-TNG & 1P\_p3\_2 & 0.004 & 2.09 & \color{red}{-3.00} & 0.0020 \\
CAMELS-TNG & 1P\_p3\_n2 & 0.123 & 5.88 & 1.38 & 0.0041 \\
CAMELS-TNG & 1P\_p4\_2 & 0.075 & 6.01 & 1.17 & 0.0025 \\
CAMELS-TNG & 1P\_p4\_n2 & 0.053 & 5.98 & 0.93 & 0.0021 \\
CAMELS-TNG & 1P\_p5\_2 & 0.025 & 7.53 & 1.11 & 0.0035 \\
CAMELS-TNG & 1P\_p5\_n2 & 0.171 & 5.89 & 1.07 & 0.0030 \\
CAMELS-TNG & 1P\_p6\_2 & 0.135 & 5.72 & 1.19 & 0.0025 \\
CAMELS-TNG & 1P\_p6\_n2 & 0.020 & 6.53 & 1.06 & 0.0016 \\
\hline
IllustrisTNG & TNG300 & 0.016 & 7.55 & 0.99 & 0.0007 \\
\hline
CAMELS-SIMBA & 1P\_p1\_0 & 1.413 & 4.13 & 1.22 & 0.0075 \\
CAMELS-SIMBA & 1P\_p3\_2 & 0.959 & 1.51 & 1.37 & 0.0063 \\
CAMELS-SIMBA & 1P\_p3\_n2 & 1.000 & 5.23 & 1.13 & 0.0116 \\
CAMELS-SIMBA & 1P\_p4\_2 & 2.161 & 3.62 & 1.31 & 0.0065 \\
CAMELS-SIMBA & 1P\_p4\_n2 & 0.867 & 4.07 & 0.81 & 0.0038 \\
CAMELS-SIMBA & 1P\_p5\_2 & 0.195 & 7.24 & 0.30 & 0.0055 \\
CAMELS-SIMBA & 1P\_p5\_n2 & 2.667 & 3.67 & 1.37 & 0.0096 \\
CAMELS-SIMBA & 1P\_p6\_2 & 3.870 & 3.28 & 1.66 & 0.0082 \\
CAMELS-SIMBA & 1P\_p6\_n2 & 0.960 & 4.57 & 0.63 & 0.0055 \\

\hline
Illustris & Illustris & 0.973 & 2.84 & 2.11 & 0.0056 \\
\hline
CAMELS-TNG cosmo & 1P\_p1\_n2 & 0.198 & 4.11 & 1.62 & 0.0126 \\
CAMELS-TNG cosmo & 1P\_p1\_n1 & 0.087 & 5.44 & 1.46 & 0.0034 \\
CAMELS-TNG cosmo & 1P\_p1\_1 & 0.033 & 6.71 & 1.07 & 0.0022 \\
CAMELS-TNG cosmo & 1P\_p1\_2 & 0.009 & 7.25 & 2.32 & 0.0018 \\
CAMELS-TNG cosmo & 1P\_p2\_n2 & 0.036 & 4.10 & 2.35 & 0.0027 \\
CAMELS-TNG cosmo & 1P\_p2\_2 & 0.023 & 7.58 & 1.40 & 0.0023 \\
CAMELS-TNG cosmo & 1P\_p7\_n2 & 0.005 & 6.41 & \color{red}{4.00} & 0.0018 \\
CAMELS-TNG cosmo & 1P\_p7\_2 & 0.083 & 5.84 & 1.33 & 0.0028 \\
\hline
\end{tabular}
\end{table}

\section{Stress test to $k=5h/$Mpc}
We further explore the applicability of the three-parameter fitting formula beyond $k=3h/$Mpc. Fig. \ref{fig:3Pk5k1345} shows its performance to $k=5h/$Mpc and demonstrates that it is still applicable. Note that $b=3.87$ is fixed. 

The performance at $k=4/h$Mpc is still good, accurate to $|\Delta S|<0.03$ except the $\Omega_m=0.1$ case. However, at $k\ga 5h/$Mpc, since  gas cooling starts to dominate over feedback, our fitting scheme no longer applies. So we recommend a conservative cut of $k=3h/$Mpc and an optimistic cut of $k=5h/$Mpc. Beyond such scale, the fitting scheme (Eq. \ref{eqn:Sfit}) must be modified to incorporate new DoFs describing gas cooling induced power spectrum enhancement. This is beyond the scope of this work.

\bfinew[width=0.49\textwidth]{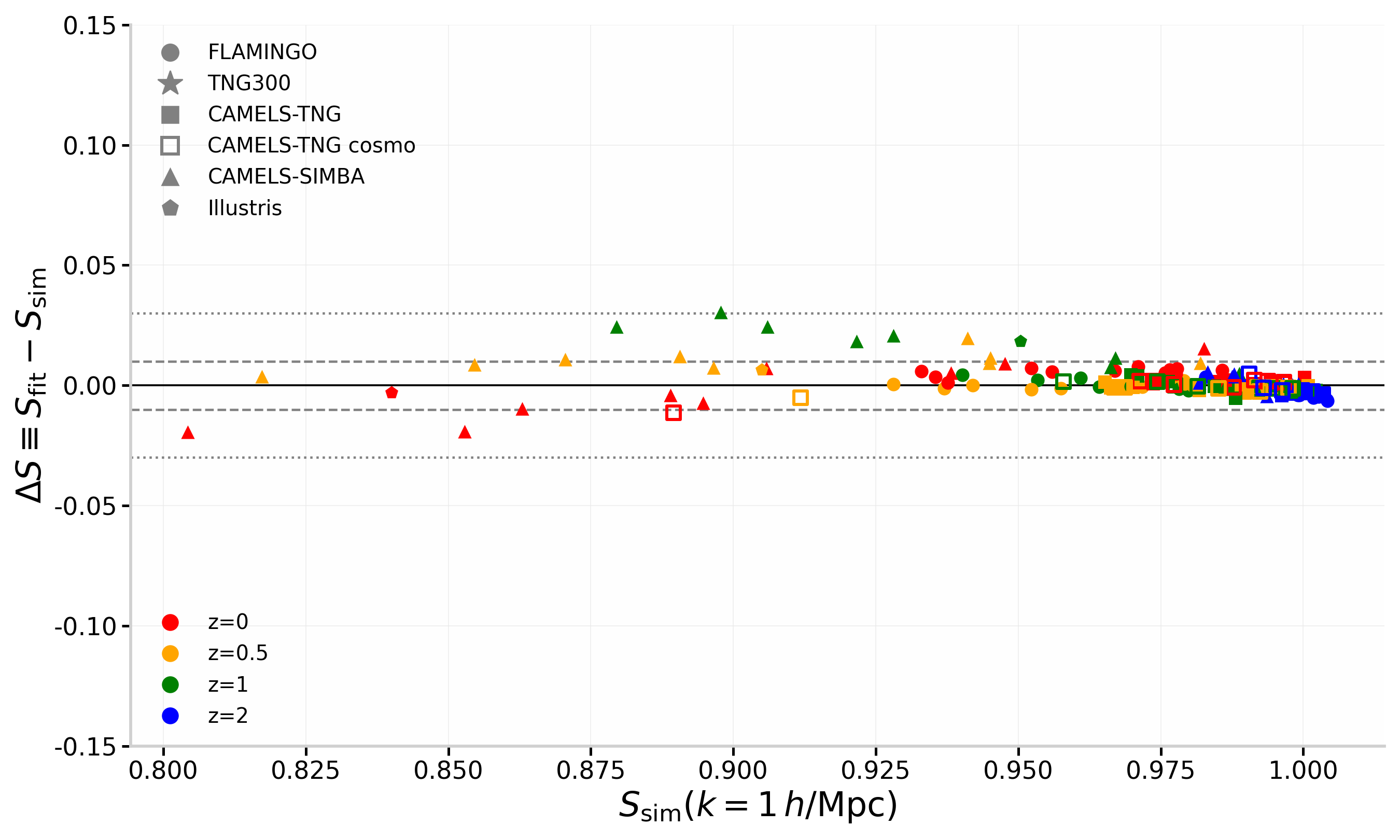}
\includegraphics[width=0.49\textwidth]{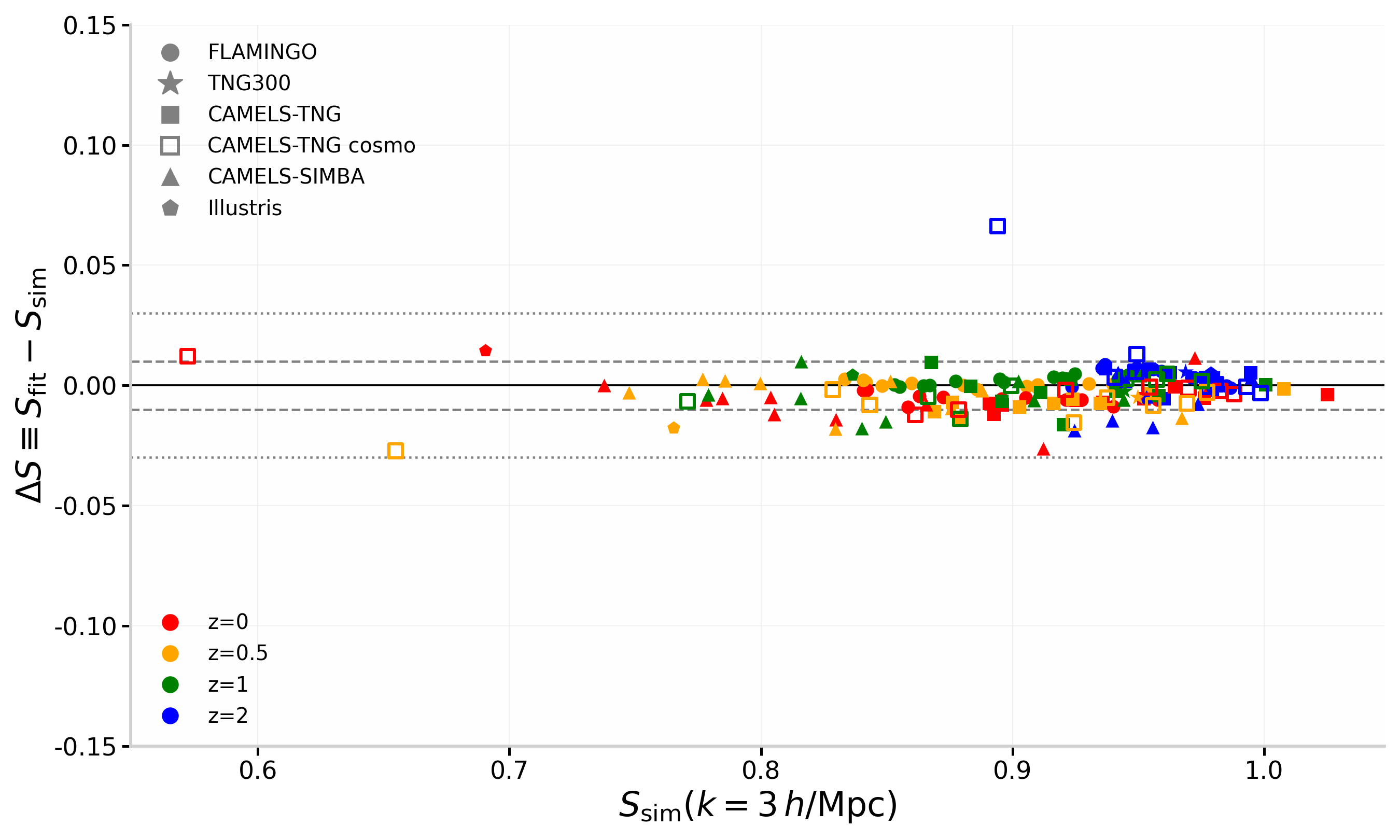}
\includegraphics[width=0.49\textwidth]{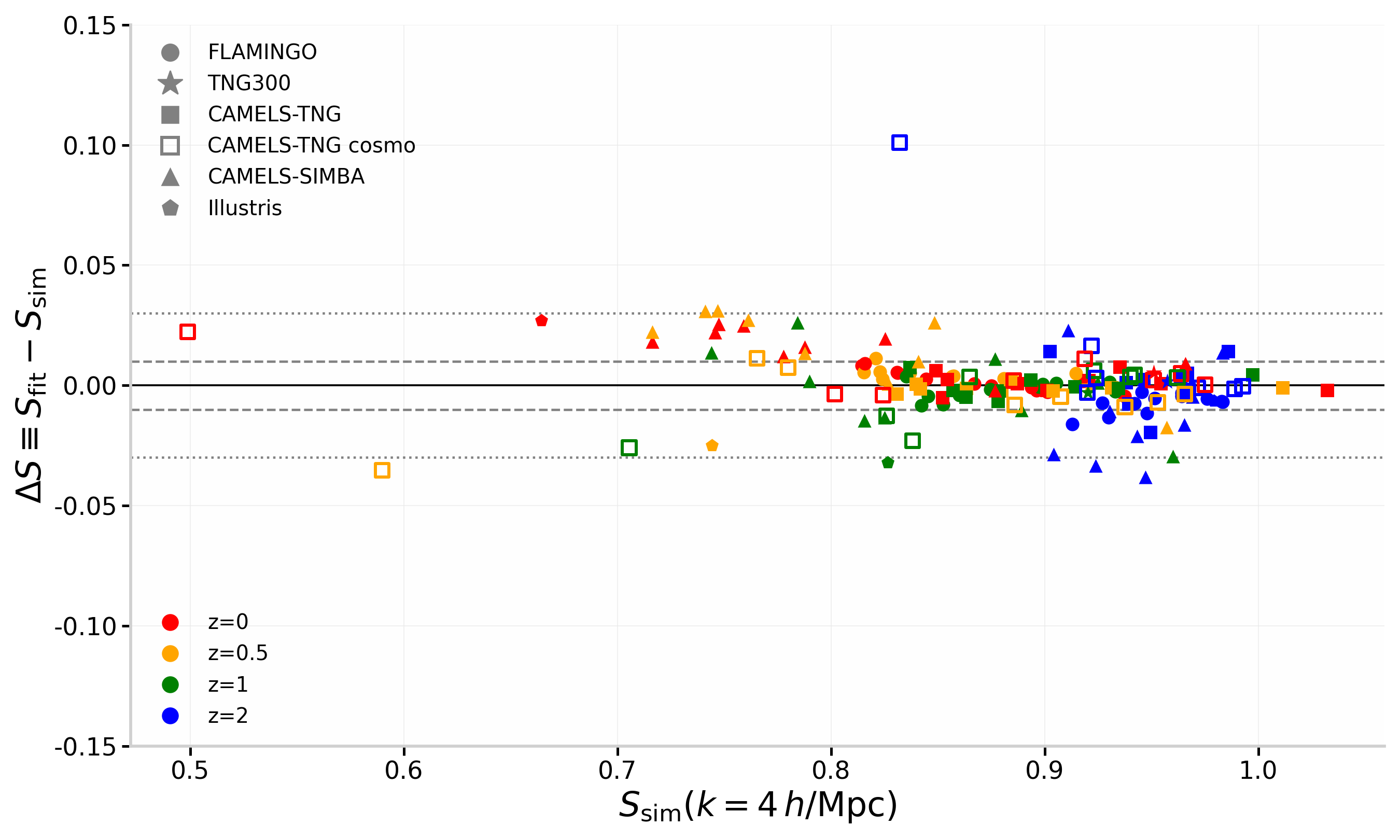}
\includegraphics[width=0.49\textwidth]{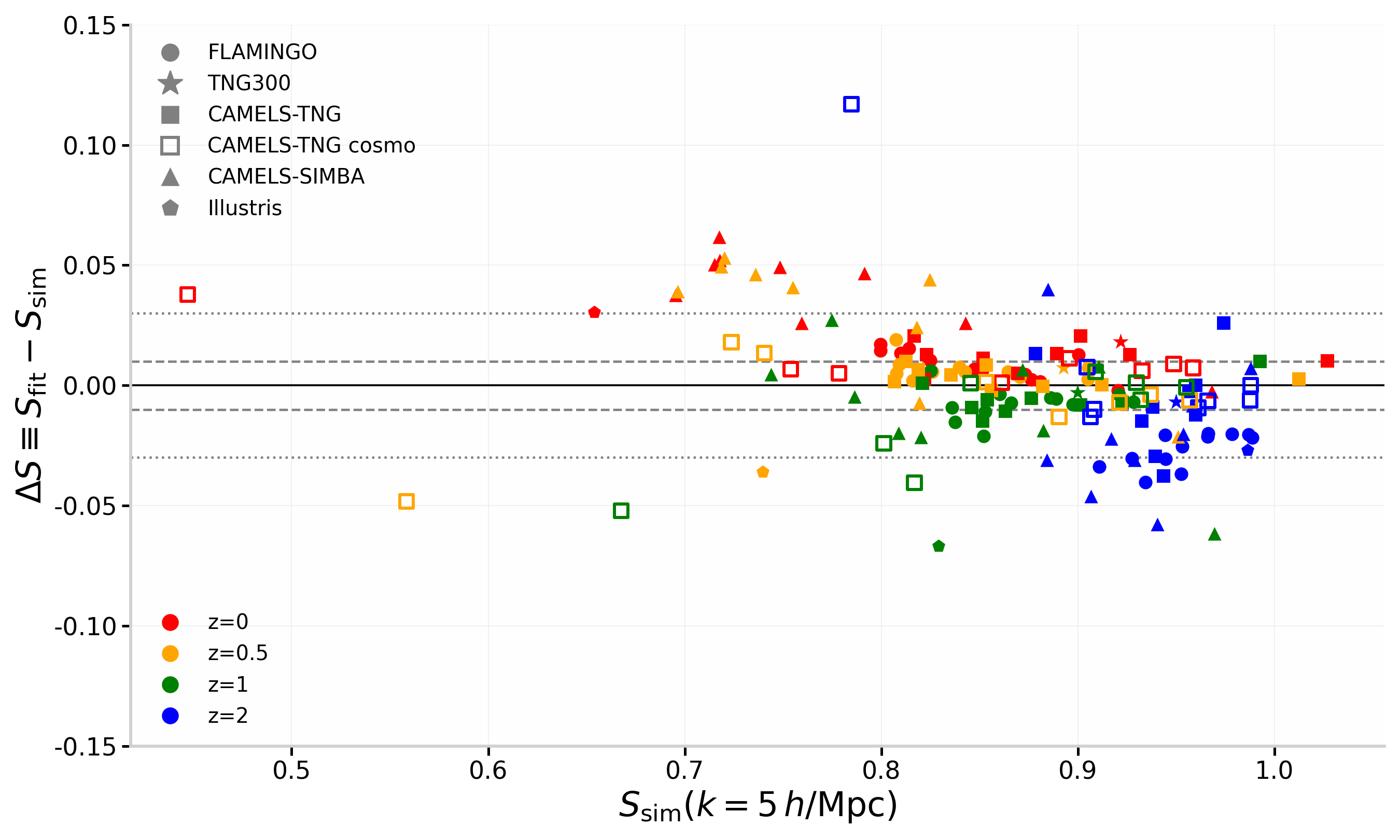}
\caption{The performance of the three-parameter formula applying to $k\leq 5h/$Mpc. At $k\ga 5h/$Mpc, gas cooling starts to dominate over feedback and enhances the matter power spectrum. This enhancement is beyond the scope of our fitting formula. A safe cut would be $k=3h/$Mpc.  \label{fig:3Pk5k1345}}
\efinew

\end{document}